\documentclass[12pt,dvips]{article}
\usepackage{rotating}
\usepackage{amsmath}
\usepackage{amsthm}
\usepackage{amsfonts}
\usepackage[dvips]{epsfig}
\usepackage{graphicx}
\usepackage{amssymb}
\usepackage{cancel}
\usepackage{pstricks} 
\usepackage{lscape}
\usepackage{color}
\usepackage{cite}
\newcommand{\beq}{\begin{equation}}
\newcommand{\eeq}{\end{equation}}
\newcommand{\ga}{\lower.7ex\hbox{$\;\stackrel{\textstyle>}{\sim}\;$}}
\newcommand{\la}{\lower.7ex\hbox{$\;\stackrel{\textstyle<}{\sim}\;$}}
\def\K{K{\"a}hler}

\begin{document}

\def\thefootnote{\fnsymbol{footnote}}

\begin{flushright}
{\tt KCL-PH-TH/2014-11}, {\tt LCTS/2014-11}, {\tt CERN-PH-TH/2014-049}  \\
{\tt ACT-3-14, UMN-TH-3332/14, FTPI-MINN-14/10} \\
\end{flushright}

\begin{center}
{\bf {\Large Resurrecting Quadratic Inflation in No-Scale \\
\vspace {0.1in}
Supergravity in Light of BICEP2}}
\end{center}

\medskip

\begin{center}{\large
{\bf John~Ellis}$^{a}$,
{\bf Marcos~A.~G.~Garc\'ia}$^{b}$,
{\bf Dimitri~V.~Nanopoulos}$^{c}$ and
{\bf Keith~A.~Olive}$^{b}$
}
\end{center}

\begin{center}
{\em $^a$Theoretical Particle Physics and Cosmology Group, Department of
  Physics, King's~College~London, London WC2R 2LS, United Kingdom;\\
Theory Division, CERN, CH-1211 Geneva 23,
  Switzerland}\\[0.2cm]
  {\em $^b$William I. Fine Theoretical Physics Institute, School of Physics and Astronomy,\\
University of Minnesota, Minneapolis, MN 55455, USA}\\[0.2cm]
{\em $^c$George P. and Cynthia W. Mitchell Institute for Fundamental Physics and Astronomy,
Texas A\&M University, College Station, TX 77843, USA;\\
Astroparticle Physics Group, Houston Advanced Research Center (HARC), Mitchell Campus, Woodlands, TX 77381, USA;\\
Academy of Athens, Division of Natural Sciences,
28 Panepistimiou Avenue, Athens 10679, Greece}\\
\end{center}

\bigskip

\centerline{\bf ABSTRACT}

\noindent  
{\small The magnitude of primordial tensor perturbations reported by the BICEP2 experiment
is consistent with simple models of chaotic inflation driven by a single scalar 
field with a power-law potential $\propto \phi^n: n \simeq 2$, in contrast 
to the WMAP and Planck results, which favored models 
resembling the Starobinsky $R + R^2$ model if running of the scalar spectral index could be neglected.
While models of inflation with a quadratic potential may be constructed in 
simple $N=1$ supergravity, these constructions are more challenging in
no-scale supergravity. We discuss here how quadratic
inflation can be accommodated within supergravity, focussing primarily on the
no-scale case. We also argue that
the quadratic inflaton may be identified with the supersymmetric partner 
of a singlet (right-handed) neutrino, whose subsequent decay could have
generated the baryon asymmetry via leptogenesis.}
 
\begin{flushleft}
March 2014
\end{flushleft}
\medskip
\noindent

\newpage

\section{Introduction}

The discovery of primordial tensor perturbations by the BICEP2 experiment~\cite{BICEP2}
would be an important step in fundamental physics, if it is confirmed, since it would prove the existence of
quantum gravitational radiation. The BICEP2 result would demonstrate simultaneously the reality of
gravitational waves, whose existence had previously only been inferred
indirectly from binary pulsars~\cite{Hulse-Taylor}, and quantization of the gravitational field.
The existence of such tensor perturbations is a generic prediction of
inflationary cosmological models~\cite{oliverev,alreview,encyclopedia}, and the BICEP2 result is strong
evidence in favour of such models, the `smoking graviton', as it were.

Moreover, different inflationary models predict different magnitudes for
the tensor perturbations, and the BICEP2 measurement~\cite{BICEP2} of the
tensor-to-scalar ratio $r$ discriminates
powerfully between models, favouring those with a large energy density
$V \sim (2 \times 10^{16}~{\rm GeV})^4$. As such, it disfavours strongly the
Starobinsky $R + R^2$ proposal~\cite{Staro,MC,Staro2} and similar models, such as Higgs inflation~\cite{HI}
and some avatars of supergravity models~\cite{ENO6,ENO7,KLno-scale,WB,FKR,fklp,AHM,pallis}. 
That said, the BICEP2 result
is in some tension with previous experiments such as the WMAP~\cite{WMAP} and
Planck satellites~\cite{Planck}, which established upper limits on $r$ and seemed to favour
very small values. We are not qualified to comment on the relative merits of
these different experiments, which may be reconciled if the scalar spectral index runs fast,
but for the purposes of this paper we take at
face value the BICEP2 measurement of $r$~\cite{BICEP2} while retaining the
measurements of the tilt in the scalar spectrum, $n_s$, found by the
previous experiments~\cite{WMAP,Planck}, with which BICEP2 is consistent.

Planck and previous experiments were in some tension with the
single-field power-law inflationary potentials of the form $\mu^{4-n} \phi^n$
where $\mu$ is a generic mass parameter.
Among models with $n \ge 2$, that might be related directly to models
with fundamental scalar fields $\phi$, models with $n = 2$ provided the
least poor fits to previous data. However, even such quadratic models were
barely compatible with the Planck results at the 95\% CL~\cite{Planck}. Quadratic
models \cite{m2} are, in some sense, the simplest, since just such a single form of 
the potential could describe dynamics throughout the inflationary epoch
and the subsequent field oscillations, unlike monomial potentials of
the form $\phi^n: n \ne 2$, which would require modification at small $\phi$
in order to accommodate a particle interpretation. Moreover, there are
motivated particle models that would yield a quadratic potential, e.g.,
for the scalar supersymmetric partner of a singlet (right-handed) neutrino
in a Type-I seesaw model of neutrino masses~\cite{ERY}. Such a model would make
direct contact with particle physics, and the decays of sneutrino inflatons
could naturally yield a cosmological baryon asymmetry via leptogenesis.
Such a scenario would be a step towards a physical model of inflation. 

In this paper we first set the scene by revisiting simple slow-roll inflationary models based on single-field
monomial potentials of the form $\mu^{4-n} \phi^n$ in light of the BICEP2 
result~\cite{BICEP2}. We derive and explore the validity of a general consistency 
condition on monomial models:
\begin{equation}
r \; = \; 8 \left( 1 - n_s - \frac{1}{N} \right) \, ,
\label{consistency}
\end{equation}
where $N$ is the number of e-folds of inflation. This consistency
condition is comfortably satisfied for the value $r = 0.16^{+0.06}_{-0.05}$
(after dust subtraction) indicated by BICEP2~\cite{BICEP2}, and the values $n_s = 0.960 \pm 0.008$ 
and $N = 50 \pm 10$ consistent with this and other experiments~\cite{WMAP,Planck}.
The consistency condition (\ref{consistency}) is independent of the
monomial power index $n$, but in the quadratic case $n = 2$ one finds
for $N = 50$ that $n_s = 0.960$ and $r = 0.16$, in perfect agreement
with the data. On the other hand, an $n = 4$ potential would have $\delta \chi^2 \sim 8$,
as we discuss later.

Global supersymmetry accommodates very naturally~\cite{ENOT} a single-field $\phi^2$
model, one example being the sneutrino model~\cite{ERY} mentioned above.
However, one should embed such a model in the framework of supergravity~\cite{SUGRA}.
The first attempt at constructing an inflationary model in $N=1$ supergravity
proposed a generic form for the superpotential for a single inflaton~\cite{nost},
the simplest example being $W = m^2 (1-\Phi)^2$~\cite{hrr}.
However, these models relied on an accidental cancellation between contributions to the inflaton mass~\cite{lw}.
Such cancellations are absent in generic supergravity models, which typically yield effective potentials with higher
powers of the inflaton field~\cite{eta,alreview,encyclopedia}. These problems can be alleviated 
either by employing a shift symmetry in the inflaton direction \cite{kyy} or through
no-scale supergravity~\cite{no-scale,LN,EENOS}. 
Since no-scale supergravity arises as the effective field theory of compactified string
theory~\cite{Witten}, and is an attractive framework for sub-Planckian physics~\cite{ENO8}, this is
an appealing route towards embedding quadratic inflation in a more complete theory.

The bulk of this paper explores possibilities for obtaining a quadratic inflaton potential in the context of supergravity.
After briefly reviewing models that invoke a shift symmetry, we turn our focus to no-scale supergravity models.
We distinguish two classes of such models, which are differentiated by how the moduli in the theory 
obtain their vevs. We give an explicit example that 
incorporates supersymmetry breaking and a simple quadratic inflationary potential embedded in no-scale
supergravity with a stabilized K\"ahler modulus.

\section{Inflation with power-law potentials}

\subsection{General power-law potentials}

We work in the slow-roll approximation~\cite{encyclopedia}, where the magnitude
of the scalar density perturbations implies that
\begin{equation}
\label{Amag} 
\left(\frac{V}{\epsilon}\right)^{\frac{1}{4}} \; =  \; 0.0275 \times M_{Pl} \, ,
\end{equation}
where $V$ is value of the effective inflationary potential and $\epsilon$ is a
slow-roll parameter given by~\cite{encyclopedia}
\begin{equation} 
\epsilon \; = \; \frac{1}{2} M_{Pl}^2 \left( \frac{V'}{V} \right)^2 \, ,
\label{epsilon}
\end{equation}
where, here and subsequently, the prime denotes a derivative with respect to the inflaton field $\phi$, and
$M_{Pl}$ corresponds to the reduced Planck mass, $2.4 \times 10^{18}$ GeV.
Other slow-roll parameters are~\cite{encyclopedia}
\begin{equation} 
\eta \; = \; M_{Pl}^2 \left( \frac{V''}{V} \right)  ;   \; \xi \; = \; M_{Pl}^4 \left( \frac{V'V'''}{V^2} \right) \, .
\label{etaxi}
\end{equation}
CMB observables can be expressed as follows in terms of the slow-roll parameters:
\begin{eqnarray}
{\rm Tensor-to-scalar~ratio}~r: & r \; = & 16 \epsilon \, , \\
{\rm Scalar~spectral~tilt}~n_s: & n_s \; = & 1 - 6 \epsilon + 2 \eta \, , \\
{\rm Running~of~scalar~index}~\alpha_s: & \alpha_s \; = & 2 \xi + 16 \, \eta\,  \epsilon - 24 \, \epsilon^2 \, .
\label{observables}
\end{eqnarray}
In addition to the above expressions, we note the formula
\begin{equation}
N \; = \; \int^{\phi_e}_{\phi_i} \left( \frac{V}{V'} \right) d\phi
\label{efolds}
\end{equation}
for the number of e-folds of inflation between the initial and final values of the inflaton field $\phi_{i,f}$.
Within this framework, the BICEP2 measurement $r = 0.16^{+0.06}_{-0.05}$~\cite{BICEP2} 
(after subtraction of an estimated dust contribution) provides a first
direct determination of $\epsilon \sim 0.01$ and hence, via (\ref{Amag}), a determination of the
potential energy density during inflation: $V \simeq (2 \times 10^{16}~{\rm GeV})^4$. The
measurement of $n_s \simeq 0.960$ then implies that also $\eta \sim 0.01$. Clearly, these
determinations are consistent with the slow-roll approximation.

As already mentioned, there is tension between the BICEP2 measurement of $r$
and the Planck upper limit, which could be alleviated if there were significant running of the
scalar index: $\alpha_s \sim - 0.02$~\cite{BICEP2}.
Since $\epsilon$ and $\eta$ are both ${\cal O}(10^{-2})$, corresponding to $V' \sim 0.1/M_{Pl}$
and $V'' \sim 0.01/M_{Pl}^2$, such a magnitude of
the scalar spectral index would require $\xi \sim 0.01$ and hence $V''' \sim 0.1/M_{Pl}^3$. 
In this case, the variation in $V''$ over a range $\Delta \phi = {\cal O}(10 M_{Pl})$ is $\Delta V'' \sim 1/M_{Pl}^2$,
which is difficult to reconcile with the estimate of $\eta$ from measurements of $r$ and $n_s$,
and indeed the slow-roll approximation in general. We therefore assume instead that the
running of the spectral index is negligible, in which case the tension between BICEP2
and Planck cannot be alleviated.

We now consider the simplest possible class of single-field models of inflation, namely a monomial
of the form $V = \mu^{4-n} \phi^n$. In this case, the slow-roll parameters have the expressions
\begin{equation}
\epsilon \; = \; \frac{n^2}{2}  \frac{M_{Pl}^2}{\phi^2}; \; \eta \; = \;  n(n-1)\frac{M_{Pl}^2}{\phi^2} \, ,
\label{monomialvalues}
\end{equation}
corresponding to
\begin{equation}
r \; = \; 8 n^2 \frac{M_{Pl}^2}{\phi^2}; \; n_s \; = \; 1 - n(n+2) \frac{M_{Pl}^2}{\phi^2} \, ,
\label{monomialrns}
\end{equation}
where we have now suppressed the suffix $i$ in $\phi_i$, and the number of e-folds is
\begin{equation}
N \; = \; \frac{1}{2n} \frac{\phi^2}{M_{Pl}^2}  \, ,
\label{monomialN}
\end{equation}
if we assume that $\phi_f \ll \phi_i = \phi$.
These expressions yield one consistency condition that is independent of $n$ and $\phi$, namely
\begin{equation}
r \; = \; 8 \left( 1 - n_s - \frac{1}{N} \right) \, ,
\label{consistency2}
\end{equation}
as noted earlier. As also noted earlier, the 68\% CL ranges indicated by 
BICEP2 and other experiments~\cite{BICEP2,WMAP,Planck},
$r = 0.16^{+0.06}_{-0.05}, n_s = 0.960 \pm 0.008$, combined with the expected number of
e-folds $N = 50 \pm 10$, satisfy comfortably the consistency relation (\ref{consistency2}).
This is not the case for the Planck upper limit on $r$ if the scalar spectral index does not run,
namely $r < 0.08$ at the 68\% CL. 

\subsection{Quadratic Inflation}
\label{sect:QI}

Given the consistency of the single-field monomial potential with experiment,
one may then ask what value of $n$ is favoured. The expressions (\ref{monomialrns}, \ref{monomialN})
can be used to derive two expressions for $n$ that are independent of $\phi$, namely
\begin{equation}
n \; = \; \frac{r N}{4} ; \; n \; = \; 2 \left[ N (1 - n_s) - 1 \right] \, ,
\label{nn}
\end{equation}
which can be combined to yield (\ref{consistency2}). Inserting $r = 0.16^{+0.06}_{-0.05}$, 
$N = 50 \pm 10$ and $n_s = 0.960 \pm 0.008$, we find the values
\begin{equation}
n \; = \; 2.0^{+0.9}_{-0.8}; \; n \; = \; 2.0 \pm 1.1 \, .
\label{nvalues}
\end{equation}
Clearly these are highly consistent with the quadratic case $n = 2$. The cases
$n = 1, 3$ ($\Delta \chi^2 \sim 2$) cannot be excluded, whereas $n = 4$ ($\Delta \chi^2 \sim 8$)
is strongly disfavoured~\footnote{Potentials with combinations of quadratic and quartic terms
have also been considered recently in light of BICEP2: see~\cite{combine}.}. 
However, since the $\phi$ and $\phi^3$ potentials are not bounded below for negative $\phi$,
they would certainly require modification in this region, as well as near $\phi = 0$ in order to
have a particle interpretation, so we disfavour them.
We are therefore led to consider quadratic inflation in more detail.

In the case $n = 2$, the analysis of~\cite{ERY} showed that mass of the inflaton, 
$m = \sqrt{2} \mu = 1.8 \times 10^{13}$~GeV
$= {\cal O}(10^{-5} M_{Pl})$,
and we see from (\ref{monomialN}) that one requires an initial field value $\phi = \sqrt{200} M_{Pl}$,
corresponding to $V = \mu^2 \phi^2 \simeq (2 \times 10^{16}~{\rm GeV})^4$. The small value of $m$ 
(or, equivalently, $\mu$) raises the usual problems of fine-tuning and naturalness in the presence of 
quadratic divergences in the quantum corrections to the effective field theory. This issue would not
arise if the inflaton $\phi$ is embedded in a supersymmetric theory. We also note that, if one relaxes
the monomial assumption, any contribution of the form $\Delta V = \lambda \phi^4$ would need to have
$\lambda \la 10^{-13}$. In a supersymmetric theory, $\lambda = 2 y^2$, where $y$ is some Yukawa
coupling. Both $\lambda$ and $y$ would receive only logarithmic wave-function renormalization,
so that small values are technically natural. Moreover, since the Yukawa
coupling of the electron $\sim 2 \times 10^{-6}$, the constraint on $\lambda$ does not seem
unreasonable in a supersymmetric model. These are among the reasons why we think that
``inflation cries out for supersymmetry''~\cite{ENOT}. Within this framework, we pointed out specifically that
suitably small values of the density perturbations could be accommodated naturally.

Supersymmetrizing the $m^2 \phi^2/2$ potential is a first step in incorporating
BICEP2-compatible inflation into a more complete physics model.
A second step is to identify the inflaton with the scalar partner of a singlet
(right-handed) neutrino in a Type-I seesaw model of neutrino masses~\cite{ERY}. In this case, the
sneutrino inflaton decays directly into Standard Model Higgs bosons and leptons, and one-loop
effects naturally generate a CP-violating lepton asymmetry. It was shown in~\cite{ERY} that there is a large
range of parameters in which sphalerons then generate an acceptable cosmological baryon asymmetry~\footnote{The
large energy density during inflation indicated by BICEP2 tends to indicate a high reheating temperature,
which would yield a high gravitino density, but this is not necessarily a problem if the gravitino mass
is high enough - a possibility compatible with the specific inflationary supergravity scenarios discussed later.}. 
Sneutrino inflation seems to us a very attractive scenario for linking early cosmology to particle
physics in a testable way. In this scenario, requiring that lepton-number violation be absent would forbid
any trilinear Yukawa interaction between neutrino superfields that could generate
a quartic sneutrino coupling $\lambda$.

Within the Type-I seesaw model one is led naturally to consider the possibility that
two or three sneutrinos might play r\^oles during the inflationary epoch~\cite{EFS}.
It was found that they could, in general, {\it decrease} $r$ compared to the single-sneutrino model.
This reduction would be accompanied by an increase in $n_s$ in a two-sneutrino
model, but not necessarily in a three-sneutrino model. These possibilities illustrate the
importance of detailed measurements of the tensor modes as well as refining the
measurement of $n_s$. A multi-sneutrino scenario could accommodate a value of $r$
intermediate between the values currently favoured by Planck and BICEP2.
Another example capable of yielding an intermediate value of $r$ is the
Wess-Zumino model~\cite{CEM}, but we do not pursue these possibilities here.

\section{Quadratic Inflation in Simple Supergravity}

The scalar potential in $N=1$ supergravity is given by 
\beq
V = e^G \left( G_i G^{i \bar j}G_{\bar j} -3 \right), 
\label{scalarpot}
\eeq
where we can write $G$ in terms of a K\"ahler potential $K$ and superpotential $W$
\beq
G = { K} + \log |{ W}|^2 \ ,
\eeq
giving
\beq
V=e^{ K}\left({ K}^{i\bar j}D_i W\bar D_{\bar j}\bar{ W}-3| W|^2\right),
\label{eqn:SUGRApotential}
\eeq
where $D_i W\equiv\partial_i W+{ K}_iW$. The first attempt at chaotic inflation in supergravity was made in \cite{gl1}.

For generic \K\, potentials, the exponential prefactor typically leads to the
$\eta$-problem.  
An elegant mechanism for avoiding the $\eta$-problem in supergravity with
canonical kinetic terms employs a shift symmetry in the
K\"ahler potential~\cite{kyy,Yamaguchi:2000vm,Davis:2008fv,klr,klor,hy,hiikm}\footnote{
For recent limits on possible departures from shift symmetry, see \cite{hy}}. 
Models of this type must incorporate at least two complex fields, three if one wants
to incorporate supersymmetry breaking~\cite{klor}.
The general form of the K\"ahler potential should be $K((\phi-\bar\phi)^2,S\bar S)$,
where the shift symmetry flattens the potential in the direction of the real part of $\phi$.
The simplest choice of K\"ahler potential is
\beq
K = -\frac{1}{2}(\phi-\bar\phi)^2+S\bar S \, ,
\label{sk}
\eeq
which can be combined with a superpotential 
  \beq
W = S f(\phi) 
\label{sw}
\eeq 
to yield a simple form for the scalar potential:
\beq
V = | f(\phi)|^2 \, .
\eeq
It is clear that taking $f(\phi) = m \phi $ leads directly to the desired quadratic potential.

However, it is not immediately apparent how to embed the shift symmetry in a more fundamental framework,
and the choice (\ref{sw}) of superpotential does not lend itself to a sneutrino interpretation of the inflaton~\footnote{For
another approach to the $\eta$-problem and sneutrino inflation in supergravity, see~\cite{Ant}.}.
We are therefore led to consider other supergravity models that can yield quadratic inflation.

\section{Quadratic Inflation in No-Scale Supergravity}

We now consider how an effective
potential of the form $m^2\phi^2/2$ could be obtained in a no-scale supergravity framework~\cite{no-scale}, which is motivated by
models of string compactification~\cite{Witten}, and is hence a step towards an
ultra-violet completion of the $m^2 \phi^2$ potential, as well as being an attractive
framework for sub-Planckian physics~\cite{LN,ENO8}.
No-scale supergravity~\cite{no-scale} incorporates an SU$(N,1)$/SU$(N)$ $\times$ U(1)
symmetry leading to a \K\, potential of the form
\beq
K \; = \; -3\ln \left(T + T^* - \frac{\phi^i \phi_i^*}{3} \right) \, ,
\eeq 
where the complex field $T$ could be identified as a generic string modulus field
that parameterizes, together with $N-1$ ``matter" fields $\phi^i$, an SU$(N,1)$ no-scale manifold~\cite{no-scale,LN}. 

It is straightforward to show that we must incorporate such matter fields and consider $N \ge 2$.
To see this, recall that the minimal no-scale SU(1, 1)/U(1) model may be written in terms of a single complex
scalar field $T$ with the K\"ahler function
\begin{equation}
K \; = \; -3 \ln ( T + T^*) \, ,
\label{K3}
\end{equation}
in which case the kinetic term becomes
\begin{eqnarray}
{\cal L}_{KE} \; = \;  \frac{3}{(T + T^*)^2} \partial_\mu T^*  \partial^\mu T \, ,
\label{no-scaleLKE}
\end{eqnarray}
and the effective potential becomes
\begin{equation}
V \; = \; \frac{{\hat V}}{(T + T^*)^2} \, : \, {\hat V} \; = \; \frac{1}{3}(T + T^*) |W_T|^2 - (W W_T^* + W^* W_T)\, .
\label{effV1}
\end{equation}
There are no polynomial forms of $W(T)$ that lead to a quadratic potential for a canonically-normalized field,
and we are led to consider $N \ge 2$ models with additional matter fields.

For our purposes here, we take $N=2$ and consider theories with just
two complex fields.  
In this case, the no-scale
K\"ahler potential may be written in the form
\begin{equation}
K \; = \; - 3 \ln \left(T + T^* - \frac{\phi \phi^*}{3} \right) \, ,
\label{K21}
\end{equation}
and the canonically-normalized fields can be taken as
$z_R = K/\sqrt{6}$, $z_I =  e^{K/3} \sqrt{3/2} (T - T^*)$, and $\Phi = e^{K/6} \phi$.

\subsection{Models with the K\"ahler Potential fixed dynamically}

Within this general framework,
one possibility is to fix the argument $z_R$ of the K\"ahler potential, in which case
the scalar potential takes a form similar to that in a globally supersymmetric model, namely
\beq
V = e^{K} |W_\Phi|^2 ,
\label{sun1}
\eeq
where $W_\Phi = d W/d\Phi$. It was assumed in~\cite{EENOS} that some high-scale dynamics fixes the value of $z_R$,
and a superpotential $W = \mu^2 (\phi - \phi^4/4)$ was used, which 
yielded a potential of the form $\mu^4 |1 - \phi^3|^2$.  This is a small-field inflation model that shares
many of the same properties as the simple $N=1$ example mentioned earlier~\cite{hrr}.
Unfortunately, both models predict $n_s = .933$ and are now excluded by the Planck and other data~\cite{Planck,BICEP2}.
We also note that an early attempt at a chaotic inflation model in no-scale supergravity was made in~\cite{gl2},
though this model suffers from an instability along the inflationary path~\cite{msy2}. 

On the other hand, a quadratic potential for the inflation is easily obtained from (\ref{sun1})
by taking $W = m \phi^2/2$, again with the assumption that there is a fixed vev for $z_R$.
A more complete model of this type was considered in~\cite{abdk2},
which relied on a stabilizing field as in (\ref{sk}) and (\ref{sw}).
This model provides for a vev for $z_R$ and leads to a quadratic potential for the inflaton.
In fact, the superpotential can be taken exactly as in (\ref{sw}), namely $f = m \phi$,
but with a \K\, potential 
\beq
K = (1 + \kappa_S |S|^2 + \kappa_\rho \rho) |S^2| - 3 \ln \rho
\eeq
where $\rho \equiv e^{-z_R/3}$. The corresponding potential has a minimum at $\rho = -3/4\kappa_\rho$.
However, all these theories contain a nearly massless field associated with $z_I$~\footnote{For related models, see \cite{msy2}.}.

\subsection{Models with the K\"ahler Potential undetermined}

Alternatively, one may leave the argument $z_R$ of the K\"ahler potential undetermined,
and consider instead the possibility that $T$ is fixed.
Returning to the no-scale form for the \K\, potential given by Eq. (\ref{K21}), it was shown
previously~\cite{ENO6} that in this case a superpotential of the form
\begin{equation}
W \; = \; m \left( \frac{\phi^2}{2}   - \lambda \frac{ \phi^3}{3\sqrt{3}} \right) 
\label{oldW}
\end{equation}
with $m \simeq 1.3 \times 10^{-5}$ from the amplitude of density fluctuations and  
$\lambda \simeq 1$ reproduces the effective potential of the Starobinsky model~\cite{Staro},
which is favoured by Planck data~\cite{Planck} but disfavoured by BICEP2~\cite{BICEP2}, under the assumptions that
some `hard' dynamics fixes the K\"ahler modulus $T$:
\begin{equation}
2 \langle {\rm Re}\, T \rangle \; = \;  c \, ; \; \langle {\rm Im}\, T \rangle \; = \; \langle {\rm Im\, \phi} \rangle \; = \; 0 \, ,
\label{fix}
\end{equation}
where we assume henceforth that $c = 1$. An example of $T$ fixing was given in~\cite{ENO7},
and we return below with other examples of such strong stabilization.
The model with the superpotential (\ref{oldW}) is
one of a class of no-scale models that yield Starobinsky-like inflationary potentials~\cite{ENO7},
but here we seek variants leading to a BICEP-2 compatible potential.

\subsubsection{Models with the Inflaton identified with the K\"ahler Modulus}

Within the $N = 2$ no-scale framework, one is free to choose either $\phi$ or the modulus $T$
as the inflaton. One example of a superpotential for the latter option is~\cite{Cecotti}
\beq
W \; = \; \sqrt{3} \, m \, \phi \, (T-1/2) \, ,
\label{Cecotti}
\eeq
where $m = 1.3 \times 10^{-5}$ as before.
It has recently been observed~\cite{FeKR} that in this model Im $T$ has a quadratic potential when
Re $T$ is fixed at the global minimum of the effective potential. Unfortunately, when Im $T \ne 0$,
as would be required during inflation, the effective potential is minimized at a different value of Re $T$,
and the BICEP2-compatibility of the model is lost~\footnote{Fixing the value of $\phi$ is also an
issue for this class of models: see~\cite{ENO7} and the discussion below.}. 

\begin{figure}[!htb]
\centering
	\scalebox{1.01}{\includegraphics[angle=0,origin=c]{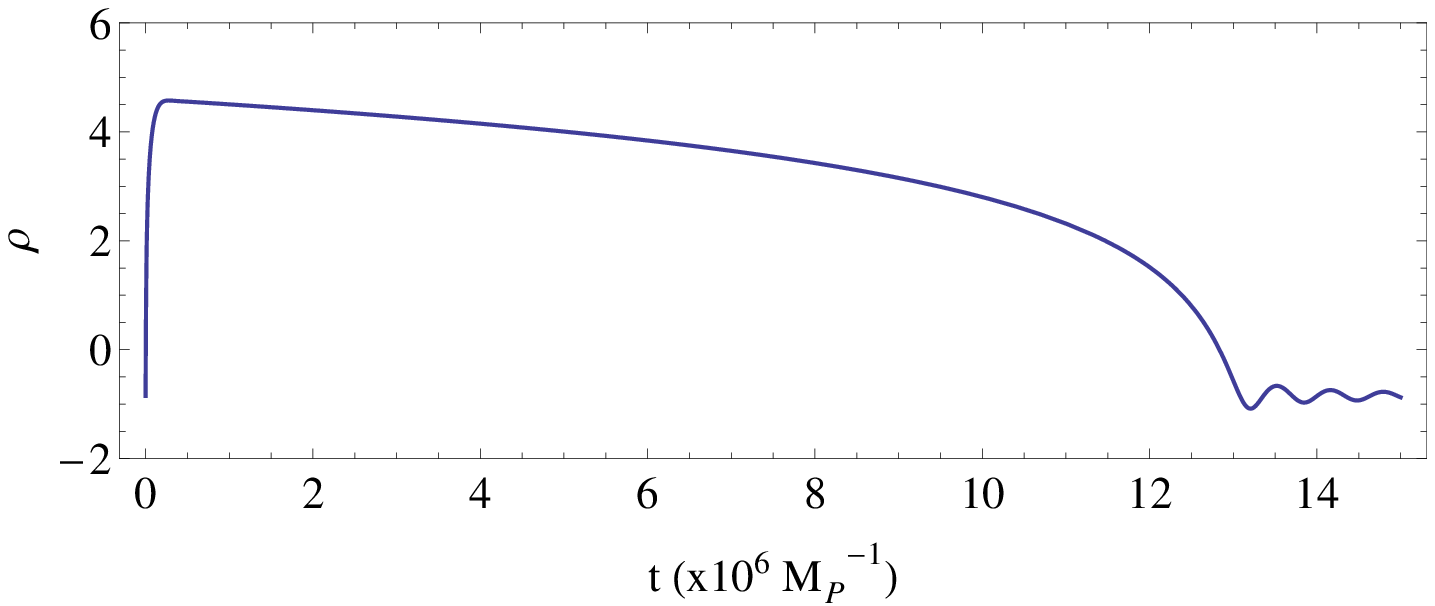}} \\
\scalebox{1.01}{\includegraphics[angle=0,origin=c]{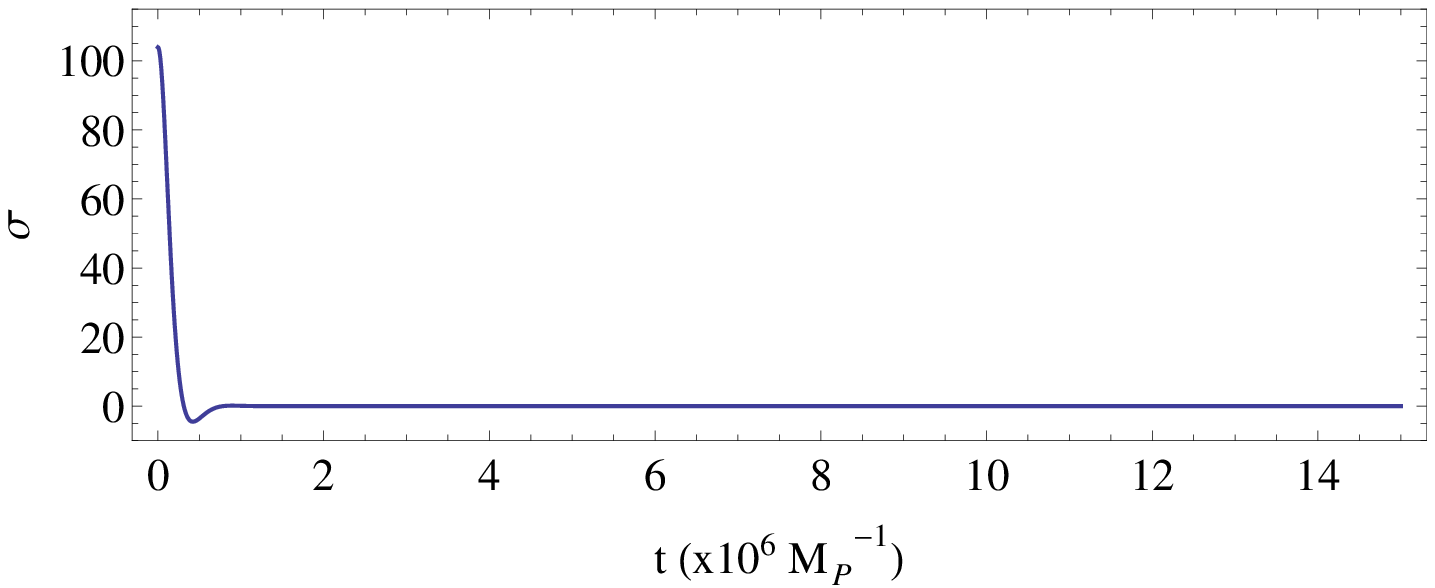}} \\
\scalebox{1.01}{\includegraphics[angle=0,origin=c]{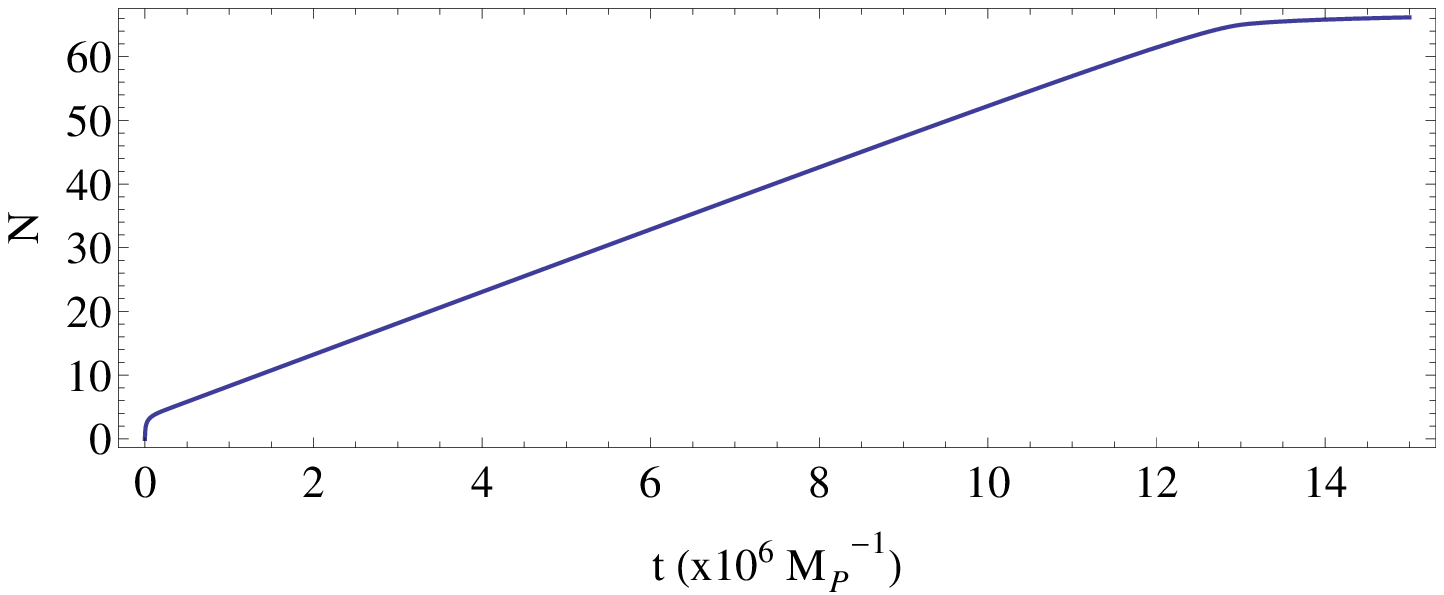}} \\
\caption{\it Analysis of the no-scale inflationary model with the inflaton identified with the K\"ahler modulus $T$
and the superpotential $W = \sqrt{3} m \phi (T-1/2)$ (\protect\ref{Cecotti}), 
assuming a suitable large initial value of ${\rm Im}\, T$.
Top panel: Time evolution of $ \rho \equiv \sqrt{3/2} \ln {\rm Re}\, T$;
middle panel: evolution of $\sigma \equiv  \sqrt{6} {\rm Im}\, T$;
bottom panel: growth of the  number of e-folds $N$ during inflation.} \label{evolveFKR}
\end{figure}

Inflationary evolution in this model is illustrated in Fig.~\ref{evolveFKR}, 
where we define
\beq
T \; \equiv \; e^{\sqrt{\frac{2}{3}}\rho}+i\frac{\sigma}{\sqrt{6}}
\label{rhosigma}
\eeq
and assume that $\rho$ is set at its 
global minimum initially, $ \rho = \sqrt{3/2} \ln (1/2)$,
but assume a large initial value of $\sigma$ and follow the evolution of
$ \rho$ and $\sigma$ during inflation. We see in the top panel that $\rho$ quickly jumps to a value $> 4$
and then decreases gradually towards zero, exhibiting small oscillations at times $> 13 \times 10^6$ in
Planck units. Conversely, we see in the middle panel that
$\sigma$ relaxes rapidly to zero, exhibiting a small overshoot at a time $\sim 0.4 \times 10^6$
in Planck units. Finally, we see in the bottom panel of Fig.~\ref{evolveFKR} that most of the inflationary
e-folds occur after $\sigma$ has settled to zero, and are driven by the roll-down of $\rho$.
In this particular example, the number of e-folds is 60, set by our choice for the  
initial value of ${\rm Im}\, T = \sigma/\sqrt{6}$. However, the inflaton
should be identified with ${\rm Re}\, T$, or equivalently $\rho$, and it would be Starobinsky-like.
We find the following values of the scalar tilt and the tensor-to-scalar ratio
\beq
(n_s,r)=
\begin{cases}
(0.9604,0.0044) \ \ {\rm for} \, N=50\\(0.9670,0.0031) \ \ {\rm for} \, N=60 \, .
\end{cases}
\label{FKRvalues}
\eeq
We conclude that this model provides a Planck/WMAP-compatible model of inflation,
but is not BICEP2-compatible. This problem of the original version of~\cite{FeKR}
was  also noted in~\cite{KLVC}.

\begin{figure}[!htb]
\vspace{-11cm}
\hspace{-3cm}
	\scalebox{1}{\includegraphics[angle=0,origin=c]{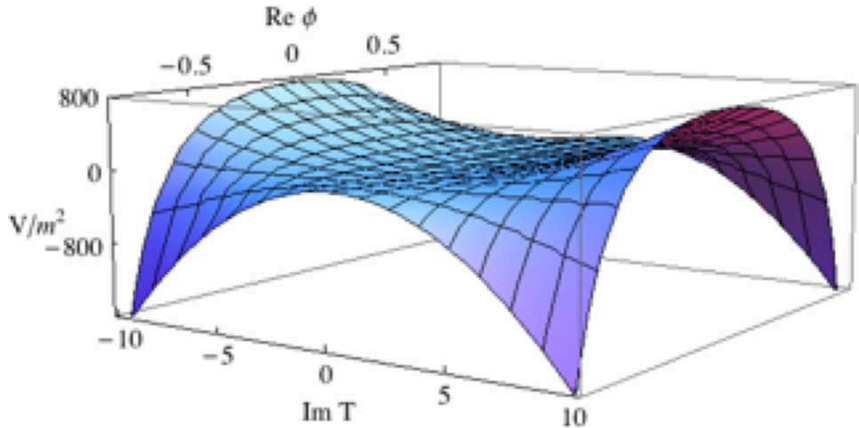}} \\
\vspace{-11cm}
\caption{\it The scalar potential of the model (\protect\ref{Cecotti}, \protect\ref{K21s1})
projected onto the $({\rm Im}\, T, {\rm Re}\, \phi)$ plane with fixed values ${\rm Re}\, T = 1/2$
and ${\rm Im}\, \phi = 0$.} \label{instab}
\end{figure}

In a revised version of~\cite{FeKR}, it was shown that the problem outlined above and in~\cite{KLVC} 
could be avoided by a modification of the \K\ potential adding a stabilization term of the type
proposed originally in~\cite{EKN3} and used more recently in~\cite{ENO7}:
\begin{equation}
K \; = \; - 3 \ln \left(T + T^* - \frac{\phi \phi^*}{3}  - \frac{(T+T*)^n}{\Lambda^2} \right)
\label{K21s1}
\end{equation}
where, as an example, the case $n=2$ and $\Lambda = \sqrt{2}$ was chosen.
The introduction of this stabilization term  leads to an acceptable potential in the ${\rm Im}\, T$ direction,
avoids the field evolution to large ${\rm Re}\, T$ in the original version of~\cite{FeKR} discussed above
and in~\cite{KLVC}, and would seem to allow for the desired quadratic inflation.  However,
the introduction of this term leads to a severe instability in the $\phi$ direction, as can be seen in Fig.~\ref{instab}
where the scalar potential is shown in the $({\rm Im}\, T$, ${\rm Re}\, \phi)$ projection for the fixed values
${\rm Re}\, T = 1/2$ and ${\rm Im}\, \phi = 0$.

\begin{figure}[!htb]
\vspace{-9cm}
\hspace{-3cm}
	\scalebox{1}{\includegraphics[angle=0,origin=c]{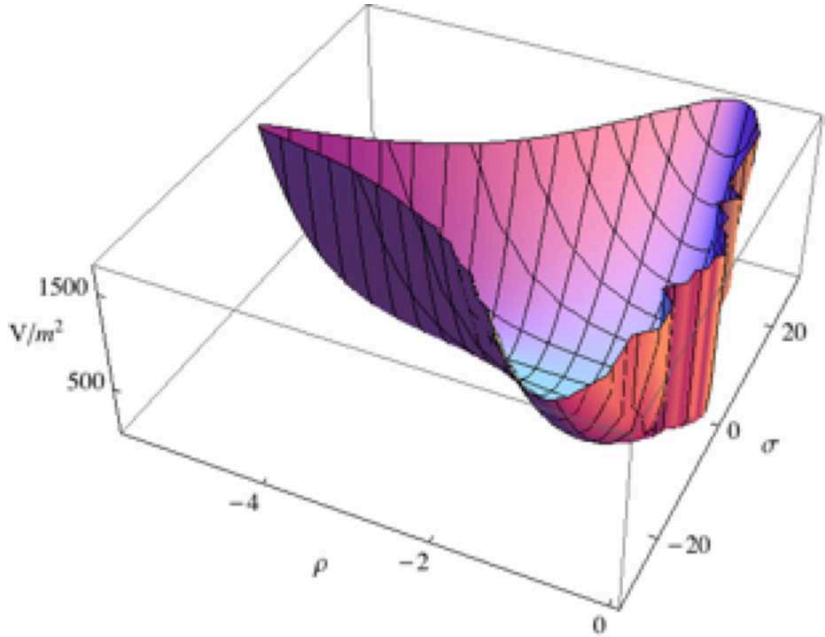}} \\
\vspace{-10cm}
\caption{\it The scalar potential of the model 
(\protect\ref{Cecotti}, \protect\ref{K21s2}) projected in the $({\rm Re}\, T$, ${\rm Im}\, T)$ plane.}
\label{retimt}
\end{figure}

\begin{figure}[!htb]
\vspace{-10cm}
\hspace{-3cm}
	\scalebox{1}{\includegraphics[angle=0,origin=c]{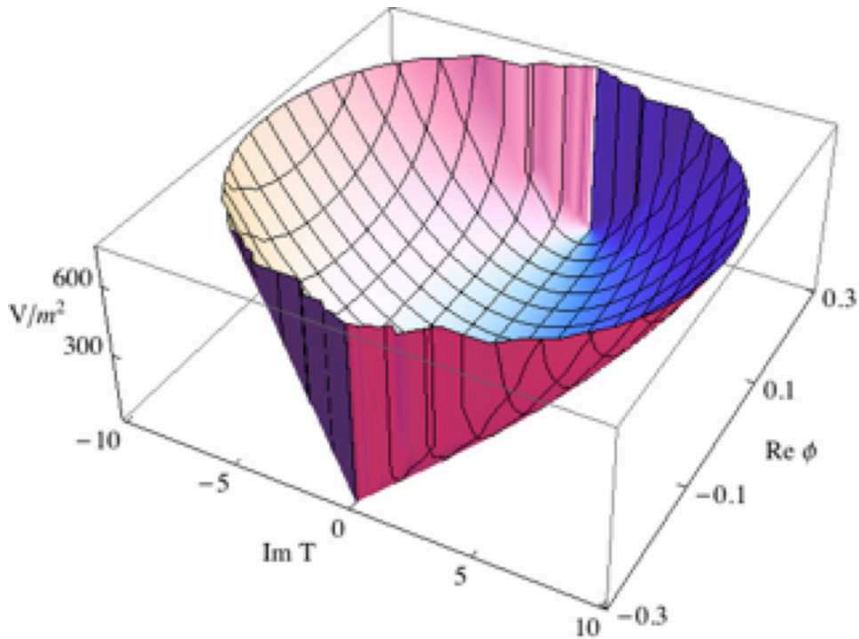}} \\
\vspace{-10cm}
\caption{\it As in Fig.~\protect\ref{instab}, but for the scalar potential of the model 
(\protect\ref{Cecotti}, \protect\ref{K21s2}).}
\label{stab}
\end{figure}

This further problem can be cured with the inclusion of a second stabilization term in the \K\ potential (\ref{K21s1}):
\begin{equation}
K \; = \; - 3 \ln \left(T + T^* - \frac{\phi \phi^*}{3}  - \frac{(T+T*)^n}{\Lambda^2} + \frac{(\phi \phi^*)^2}{\Lambda_\phi^2}  \right) \, .
\label{K21s2}
\end{equation}
where it is sufficient to take $\Lambda_\phi =  1$. The presence of the quartic term in $\phi$ in $K$,
forces $\phi$ to 0~\cite{klr,klor} and implements finally the desired quadratic inflation. The scalar potential
of the model (\ref{Cecotti}, \ref{K21s2}) at $\phi = 0$ is given by \cite{FeKR}
\beq
V = e^{-2 \sqrt{2/3} \rho} m^2 \Lambda^4 \left(
\frac{2 \sigma^2 + 3 (1-2e^{ \sqrt{2/3} \rho})^2}{16(2 e^{\sqrt{2/3} \rho} - \Lambda^2)^2} \right) \, .
\eeq
Its projection in the $({\rm Re}\, T$, ${\rm Im}\, T)$ plane is shown in Fig.~\ref{retimt},
and the $({\rm Im}\, T$, ${\rm Re}\, \phi)$ projection for
${\rm Re}\, T = 1/2$ and ${\rm Im}\, \phi = 0$ is shown in Fig.~\ref{stab}.
We note that a quadratic potential for $\sigma$ results only when $\rho$ is fixed. Fortunately,
at large $\sigma$, $\rho$ is driven to a $\sigma$-independent minimum at
$\rho = \sqrt{3/2}\ln(\Lambda^2/4)$.

\begin{figure}[h!]
\centering
\vspace{-1cm}
	\scalebox{0.55}{\includegraphics{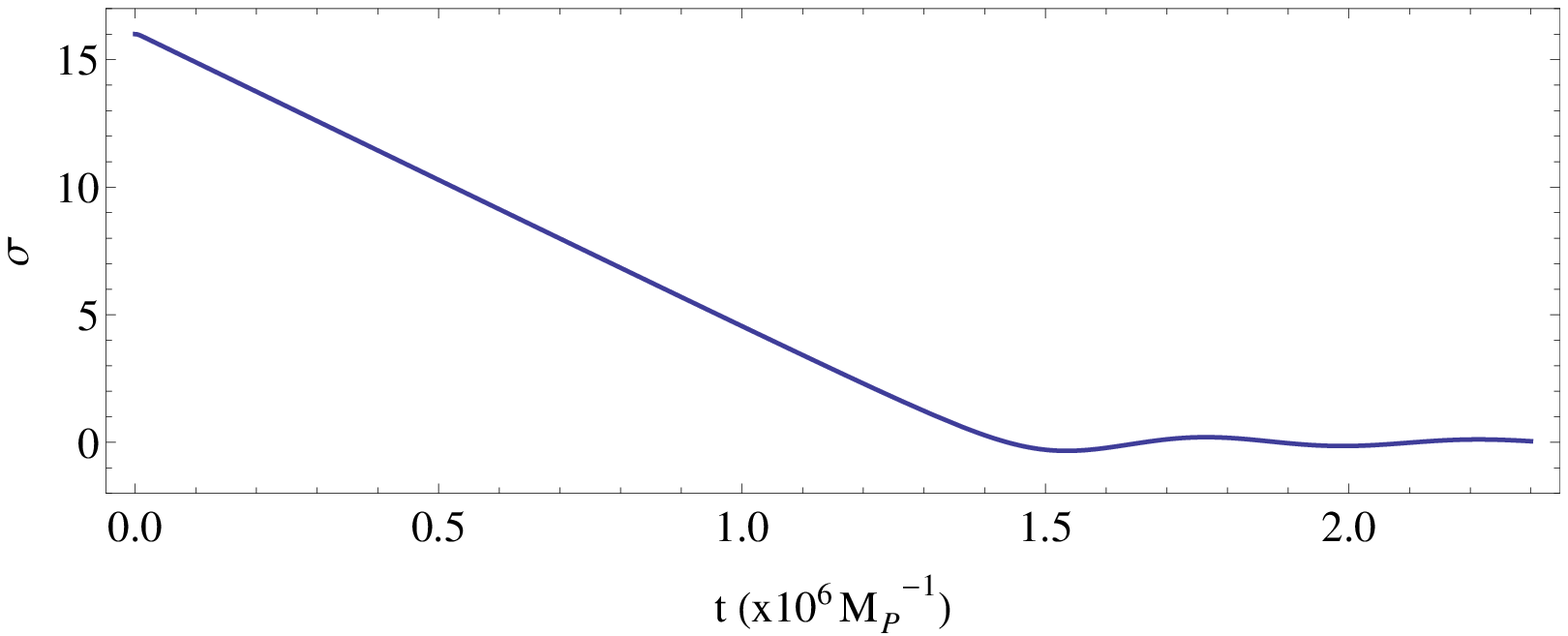}} \\
\scalebox{0.56}{\includegraphics{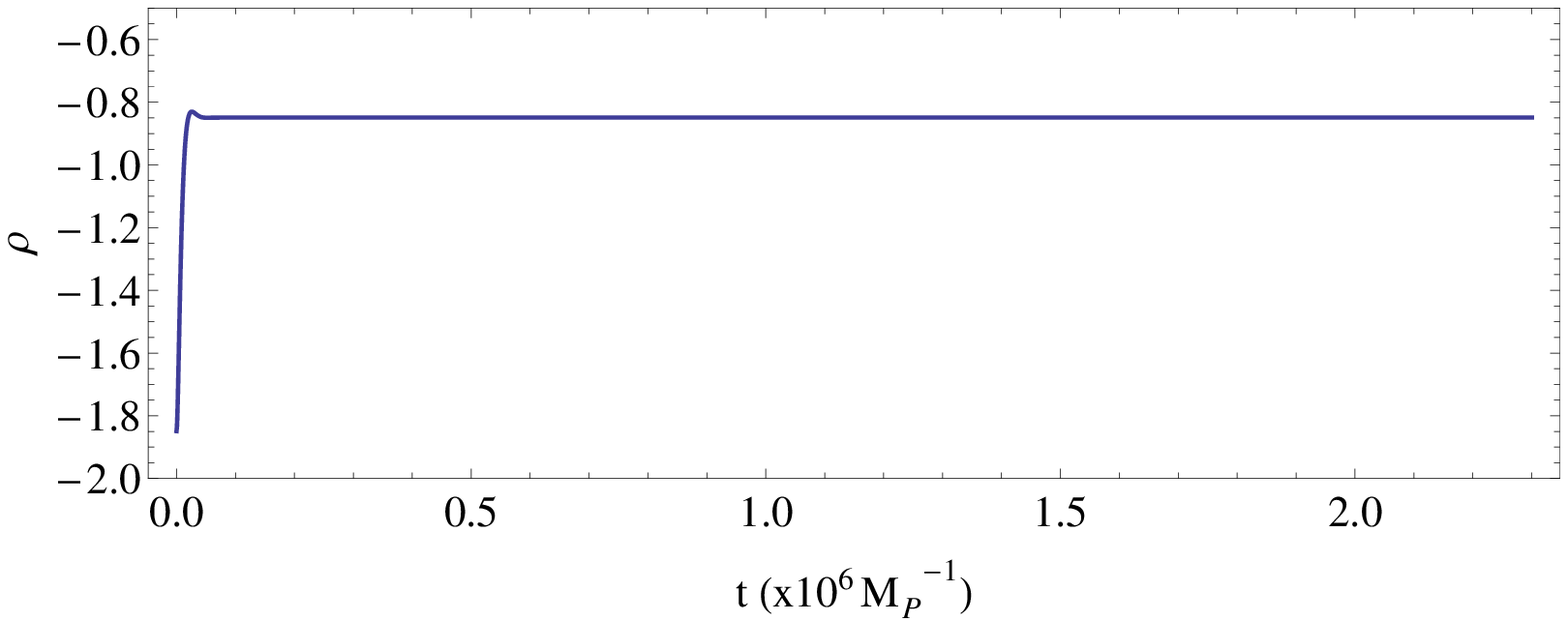}} \\
\scalebox{0.56}{\includegraphics{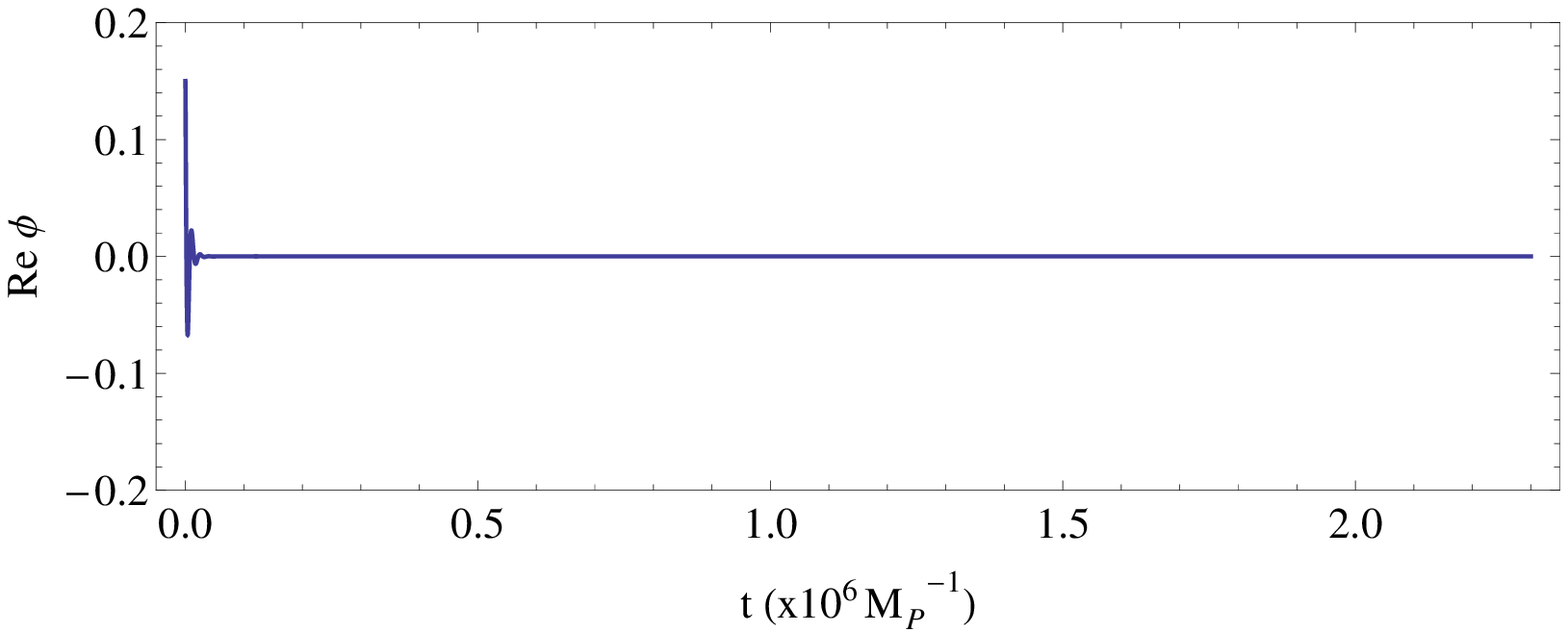}} \\
\scalebox{0.56}{\includegraphics{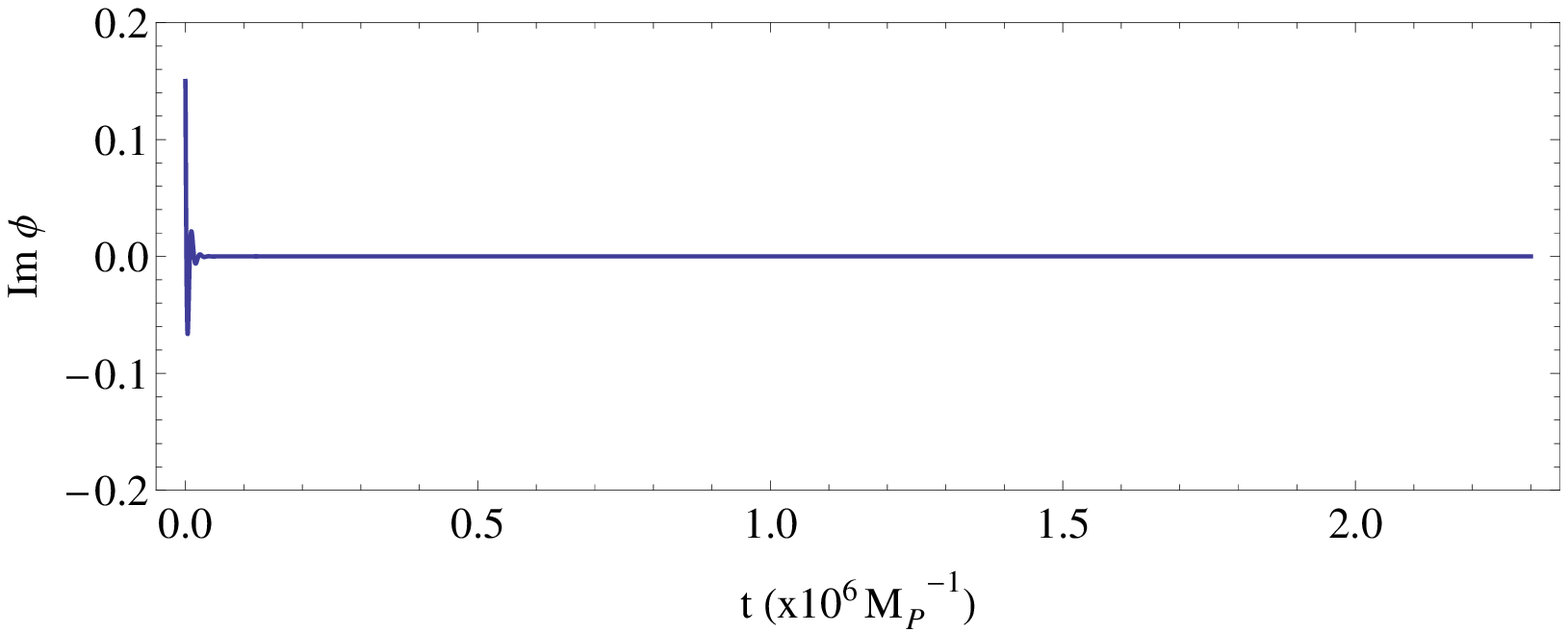}} \\
\scalebox{0.56}{\includegraphics{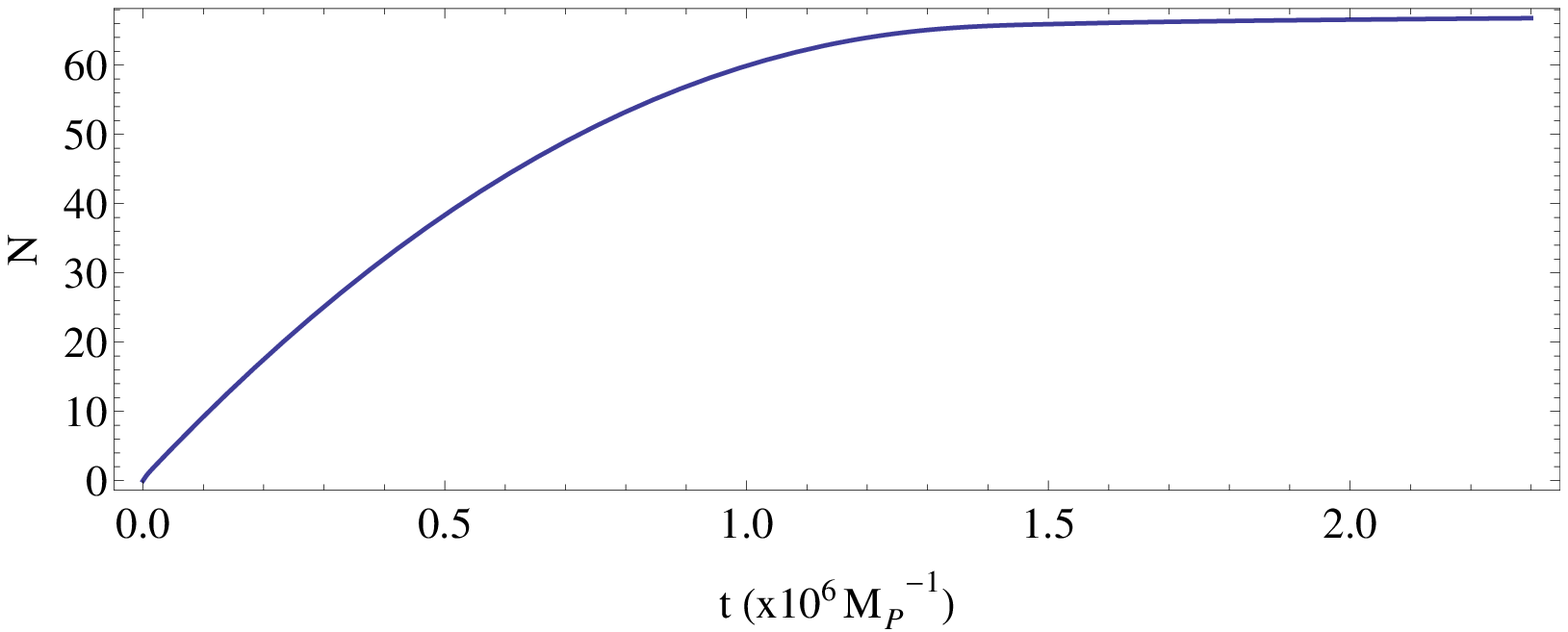}} \\
\caption{\it Analysis of the no-scale quadratic inflationary model given by the K\"ahler potential
(\protect\ref{K21s2}) and the superpotential (\protect\ref{Cecotti}).
Top panel: Time evolution of the inflaton $\sigma$, which is identified with ${\rm Im}\, T$;
second panel: evolution of $\rho$, which is identified with ${\rm Re}\, T$;
third panel: evolution of ${\rm Re}\, \phi$;
fourth panel: evolution of ${\rm Im}\, \phi$;
bottom panel: growth of the  number of e-folds $N$ during inflation.} \label{FKRevolve}
\end{figure}

We display in Fig.~\ref{FKRevolve} the evolutions of the four field components of
the model (\ref{Cecotti}, \ref{K21s2}) during inflation. The normalization of the inflaton field
$\sigma$ defined in (\ref{rhosigma}) differs from the canonical value by a numerical factor that is dependent on $\Lambda$, 
as seen in its effective Lagrangian:
\beq
{\cal L} \; = \; \left( \frac{(2-2\Lambda^2+\Lambda^4)}{2(\Lambda^2-1)^2} \right) (\partial_\mu \sigma)^2 - \left( \frac{\Lambda^4 m^2}{2(\Lambda^2-1)^2} \right) \sigma^2 \, .
\label{noncannorm}
\eeq
We note that at $\phi = 0$ the coefficient of the kinetic term for $\phi$ is proportional to $\Lambda^2/(\Lambda^2-1)$ and thus the normalization of the kinetic term is positive for $\Lambda > 1$.
Because of the non-canonical normalization, the initial value of $\sigma$ in Fig.~\ref{FKRevolve} 
must be larger than 15 in order to obtain $\sim 60$ e-folds.
We also see in (\ref{noncannorm}) that $m$ is related to the inflaton mass by a $\Lambda$-dependent numerical factor.
In the top panel of Fig.~\ref{FKRevolve} we see that the inflaton $\sigma$ falls smoothly towards zero
and then exhibits characteristic oscillations. It is crucial that $\rho$ remain relatively fixed during the inflationary evolution 
so that the $\sigma$ is driven by a quadratic potential.  The second panel shows the evolution of $\rho$,
which is related in (\ref{rhosigma}) to ${\rm Re}\, T$: it moves to its minimum at large $\sigma$ and then 
begins oscillations, but does not
modify the inflationary behaviour in an important way. For the choice $\Lambda = \sqrt{2}$, the minimum at large $\sigma$ coincides with that at $\sigma = 0$.  The qualitative behavior of the solutions will not be affected by other choices of $\Lambda$. The next two panels show the
evolutions of ${\rm Re}\, \phi$ and ${\rm Im}\, \phi$: they exhibit some damped oscillations
before relaxing rapidly to zero. Finally, the bottom panel of Fig.~\ref{FKRevolve} shows the
growth of the number of e-folds $N$ during inflation.
We find for the model (\ref{Cecotti}, \ref{K21s2}) the following values of the scalar tilt and the tensor-to-scalar ratio
\beq
(n_s,r)=
\begin{cases}
(0.9596,0.1617) \ \ {\rm for} \, N=50\\(0.9664,0.1346) \ \ {\rm for} \, N=60 \, .
\end{cases}
\label{modFKRvalues}
\eeq
We conclude that the model (\ref{Cecotti}, \ref{K21s2}) provides a satisfactory
BICEP2-compatible model of inflation.

As a BICEP2-compatible alternative, we consider the following superpotential:
\beq
W \; = \; \sqrt{3} \, m \, \phi \, T \, \ln(2T) \, .
\label{MarcosW}
\eeq
Since we seek to identify the inflaton with a component of the modulus field $T$,
we must postulate some suitable `hard' dynamics to fix $\phi$. We consider for
this purpose a modification of the K\"ahler potential that is higher order in $\phi$ and similar to that
proposed in~\cite{EKN3,ENO7,KLno-scale}:
\beq
K \; = \; -3\ln\left(T+T^*-\frac{|\phi|^2}{3} + \frac{|\phi|^4}{\Lambda^2}\right) \, .
\label{MarcosK}
\eeq
In this model the canonically-normalized inflaton field $\chi$ is given by
\beq
\chi \; \equiv \; \frac{\sqrt{3}}{2} \ln (2 T) \, ,
\label{Marcosinflaton}
\eeq
and it is easy to verify that the parameter $m$ in (\ref{MarcosW}) can be identified as the mass of the inflaton.
Indeed, at the global minimum of the effective scalar potential, the mass of the $\phi$ field is also $m$.

\begin{figure}[!h]
\centering
\vspace{-2cm}
	\scalebox{0.5}{\includegraphics{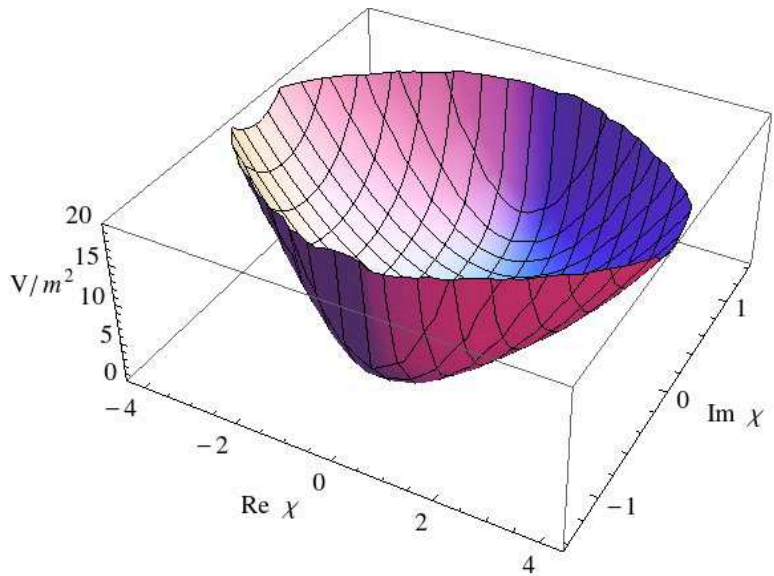}} 
\vspace{-3cm}
\caption{\it The effective potential for the real and imaginary components of $\chi$ (\protect\ref{Marcosinflaton})
in the model given by the superpotential (\protect\ref{MarcosW}),
assuming $\phi$ is fixed by (\protect\ref{MarcosK}).}
\label{Tinflation1}
\end{figure}

\begin{figure}[!h]
\centering
\vspace{-2cm}
	\scalebox{0.5}{\includegraphics{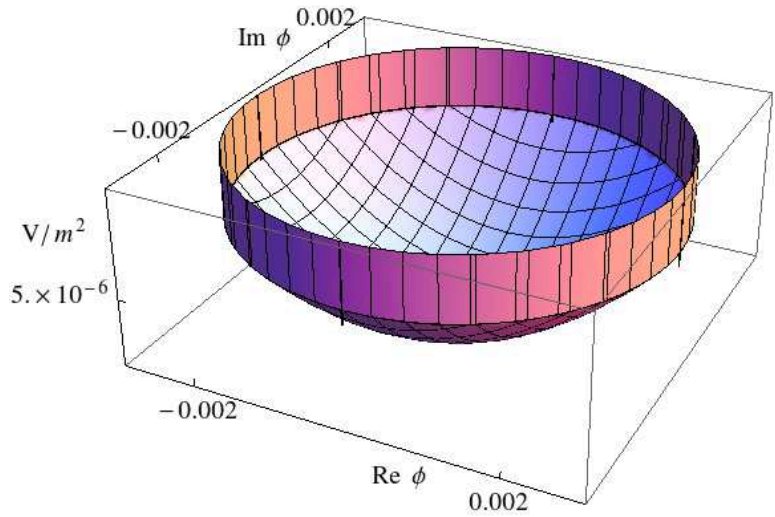}} 
\vspace{-3cm}
\caption{\it The effective potential for the real and imaginary parts of $\phi$ in the same model
(\protect\ref{MarcosW}, \protect\ref{MarcosK}, \protect\ref{Marcosinflaton}) as in Fig.~\protect\ref{Tinflation1}.}
\label{Tinflation2}
\end{figure}

\begin{figure}[!h]
\centering
\vspace{-2cm}
	\scalebox{0.5}{\includegraphics{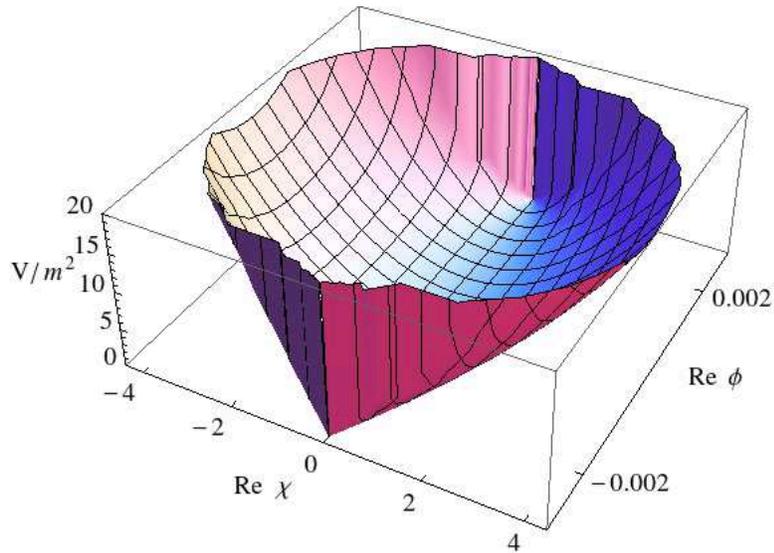}} 
\vspace{-3cm}
\caption{\it The effective potential for the real parts of $\chi$ and $\phi$ in the same model
(\protect\ref{MarcosW}, \protect\ref{MarcosK}, \protect\ref{Marcosinflaton}) as in Fig.~\protect\ref{Tinflation1}.}
\label{Tinflation3}
\end{figure}

We display the effective scalar potential of the model (\ref{MarcosW}, \ref{MarcosK}, \ref{Marcosinflaton}) in 
various projections in Fig.~\ref{Tinflation1}, \ref{Tinflation2} and \ref{Tinflation3}. Fig.~\ref{Tinflation1}
shows the effective potential for the real and imaginary components of $\chi$.
We see that both are stabilized around $\chi = 0$ and, as already mentioned, 
the effective potential for $\chi$ has a BICEP2-compatible quadratic form.
Fig.~\ref{Tinflation2} shows that the modification (\ref{MarcosK}) of the K\"ahler potential
indeed fixes both components of $\phi$. The range of $|\phi|$ is restricted by a
singularity that appears as a near-vertical wall in Fig.~\ref{Tinflation2}. Finally, Fig.~\ref{Tinflation3} shows
the effective potential for the real parts of $\chi$ and $\phi$. 
We conclude that this model provides a BICEP2-compatible model of inflation.

\subsubsection{A Model with the K\"ahler Modulus fixed dynamically}

As an alternative, we now investigate a model with the K\"ahler Modulus $T$ fixed dynamically and the
inflaton identified with the other no-scale field, using a different choice of
superpotential that yields an effective quadratic potential.

In such a no-scale scenario with $T$ fixed, the canonically-normalized inflaton field $\chi$ is defined by~\cite{ENO6}
\begin{equation}
\chi \; \equiv \; \sqrt{3} \tanh^{-1} \left( \frac{\phi}{\sqrt{3}} \right) \, ,
\label{chi}
\end{equation}
and the effective potential is
\begin{eqnarray}
\label{L2}
V & = & \frac{|W_\phi|^2}{\left[1 - \tanh \left(\frac{\chi}{\sqrt{3}} \right) \tanh \left(\frac{\chi^*}{\sqrt{3}} \right)\right]^2}\\
\label{L3}
& = & {\rm sech}^2 \left( \frac{\chi - \chi^*}{\sqrt{3}} \right) \cosh^4 \left( \frac{\chi}{\sqrt{3}} \right)
\cosh^4 \left( \frac{\chi^*}{\sqrt{3}} \right) |W_\chi|^2 \, .
\label{Vchi}
\end{eqnarray}
Let us assume that inflation occurs along the real direction, $\bar{\chi}=\chi$.
In the case of a quadratic potential, $N>50$ if $\chi\gtrsim11$. However, we see from (\ref{L2}) and (\ref{L3}) that,
for a generic superpotential, the scalar potential grows exponentially fast at large $\chi$:
\beq
V\simeq \frac{1}{16}e^{4\chi/\sqrt{3}}|W_{\phi}|^2\simeq \frac{1}{256}e^{8\chi/\sqrt{3}}|W_{\chi}|^2.
\eeq
The presence of the exponential is directly related to the presence of poles at $\phi=\pm\sqrt{3}$, since for
$|\chi| \to \infty$, $|\phi|\rightarrow\sqrt{3}$. Therefore, if we are to have large-field inflation, 
one or both of the poles must be removed: $W_{\phi}\propto(1\pm\phi/\sqrt{3})$. 
However, if this is the case, large $\chi$ implies $|\phi|\rightarrow\sqrt{3}$, and for a polynomial superpotential
$W=a\phi^n+\cdots$, $V\rightarrow  const.$, corresponding to an asymptotically scale-invariant potential 
along the inflationary trajectory, more akin to the Starobinsky scenario than to the quadratic case.

It is possible to construct a quadratic potential if one relaxes the assumption for $W$ by
allowing a non-polynomial form. Indeed, the choice
\begin{equation}
W(\phi) \; = \; \frac{m}{18} \left[ 9 - 3 \phi^2 - 2\sqrt{3} \phi (-9 + \phi^2) \tanh^{-1} \left(\frac{\phi}{\sqrt{3}} \right) + 18 \ln \left(1 - \frac{\phi^2}{3} \right) \right]
\label{funnyW}
\end{equation}
yields the effective potential $m^2({\rm Re}\,\chi)^2$,
and it is clearly possible to construct alternative
models that yield smaller values of $r$. We note that the choice (\ref{funnyW}) has a $Z_2$
symmetry: $\phi \to - \phi$, consistent with the identification of the scalar component of $\phi$
as a sneutrino.

We also note that the imaginary direction of $\phi$ cannot support inflation for a general superpotential,
due to the  presence of singularities at Im $\chi=\pm\frac{\sqrt{3}}{4}\pi$.  With the singularities removed, 
a superpotential that is polynomial in $\phi$ would result in a potential $V\left(\tan({\rm Im}\,\chi/\sqrt{3})\right)$, 
with a range limited to Im $|\chi| \le \frac{\sqrt{3}}{2}\pi$.

\subsection{A Modified No-Scale Model}

As an alternative, one may consider the modified no-scale K\"ahler potential~\footnote{Such a form could
appear if $\phi$ lies in a different modular sector, with the other modulus fixed by dynamics that we do discuss here.}
\begin{equation}
K \; = \;  -3 \ln \left( T + T^* \right) + |\phi|^2 \, .
\label{modK}
\end{equation}
The scalar field $\phi$ is now canonical, and 
in this case the scalar potential is of the form
\begin{equation}
V \; = \; e^{|\phi|^2} \left[|\phi|^2 |W|^2 + |W_\phi|^2 + (\phi W_\phi + h.c.) \right] \, ,
\label{modV}
\end{equation}
assuming that the superpotential is a function of $\phi$ only and 
where we have again below set $c = 1$ (see below for a mechanism which accomplishes this).
It is then easy to see that the choice
\begin{equation}
W \; = \; e^\frac{-\phi^2}{2} \left(\tilde{m} - \frac{m}{2} \phi^2 \right)
\label{modW}
\end{equation}
again yields the effective potential $m^2 x^2/2$, $Re \phi = x/\sqrt{2}$. In Eq. (\ref{modW}), the presence of the 
constant $\tilde{m} $ accounts for supersymmetry breaking with the gravitino mass given by $\tilde{m} $.  Therefore, we expect
$\tilde{m}  \ll m$. 
The superpotential (\ref{modW})
also has the $Z_2$ symmetry: $\phi \to - \phi$, and is far
simpler than the previous case (\ref{funnyW}), so we select it for more detailed study.

The model (\ref{modK}, \ref{modW}) has two complex fields and hence
four degrees of freedom. In order to show that this is a satisfactory model 
of inflation, one should demonstrate that the other degrees of freedom
do not `misbehave' while the real part of $\phi$ is driving inflation.
We note first that the potential (\ref{modV}) given by (\ref{modK}, \ref{modW}) is proportional to $e^{-(\phi-\phi^*)^2/2}$.
Thus the potential rises exponentially along the Im $\phi$ direction, so that
direction is automatically stabilized. In contrast, 
the potential given by (\ref{modK}, \ref{modW}) is flat
in the directions corresponding to the real and imaginary parts of $T$,
which must be stabilized in order to obtain suitable inflation. 
This can be achieved by modifying the K\"ahler potential to become \cite{EKN3,ENO7}
\beq
K \; = \; -3\log\left(T+\bar{T}+\frac{(T+\bar{T}-1)^4+d(T-\bar{T})^4}{\Lambda^2}\right)+|\phi|^2 \, ,
\label{stable}
\eeq
in which the quartic terms in the argument of the logarithm fix
the vevs: $\langle 2 {\rm Re}\, T \rangle = 1$ and $\langle {\rm Im}\, T \rangle = 0$, providing the necessary stabilization.
The masses of the real and imaginary parts of $T$ are both given by $ 12 \tilde{m} /\Lambda$ and thus are 
hierarchically larger than the gravitino mass. This type of hierarchy was recently shown 
to be compatible with preserving the baryon asymmetry while not over-producing the dark matter
density through moduli and gravitino decays \cite{ego}. 

\begin{figure}[!h]
\centering
\vspace{-2cm}
	\scalebox{0.5}{\includegraphics{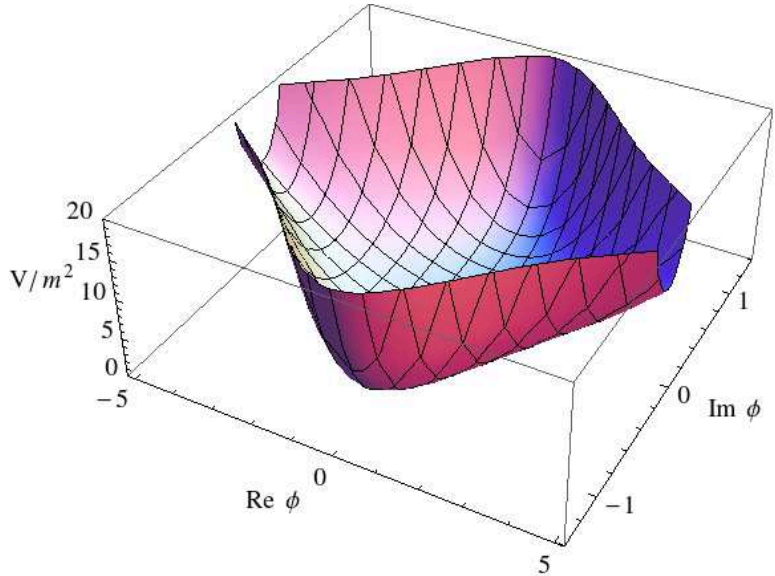}} 
\vspace{-3cm}
\caption{\it The effective potential for the real and imaginary components of $\phi$
in the model (\protect\ref{modW}, \protect\ref{stable}),
for fixed $T=1$.}
\label{pot1}
\end{figure}

\begin{figure}[!h]
\centering
\vspace{-2cm}
	\scalebox{0.5}{\includegraphics{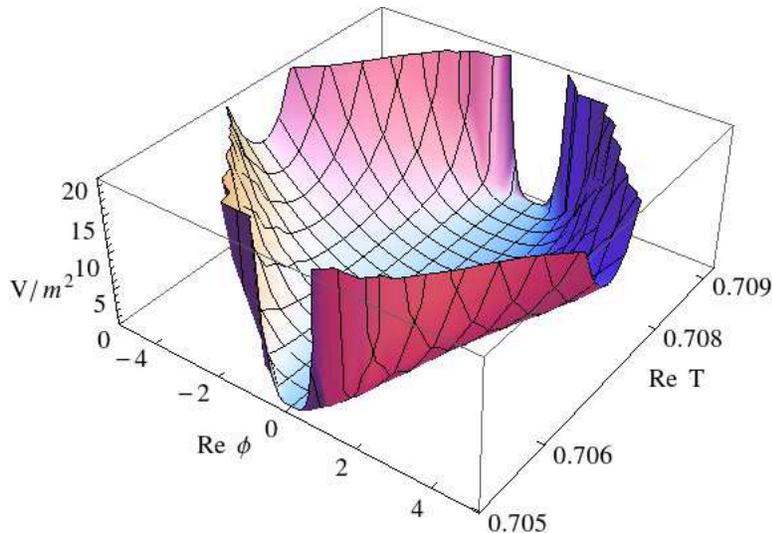}} 
\vspace{-3cm}
\caption{\it The effective potential for the real parts of $\phi$ and $T$
in the model (\protect\ref{modW}, \protect\ref{stable}), assuming that $\tilde{m}  =10^{-13}$
and $\Lambda=10^{-2}$, in the case that the imaginary parts of $\phi$ and $T$ are set to zero.}
\label{pot2}
\end{figure}

\begin{figure}[!h]
\centering
\vspace{-2cm}
	\scalebox{0.5}{\includegraphics{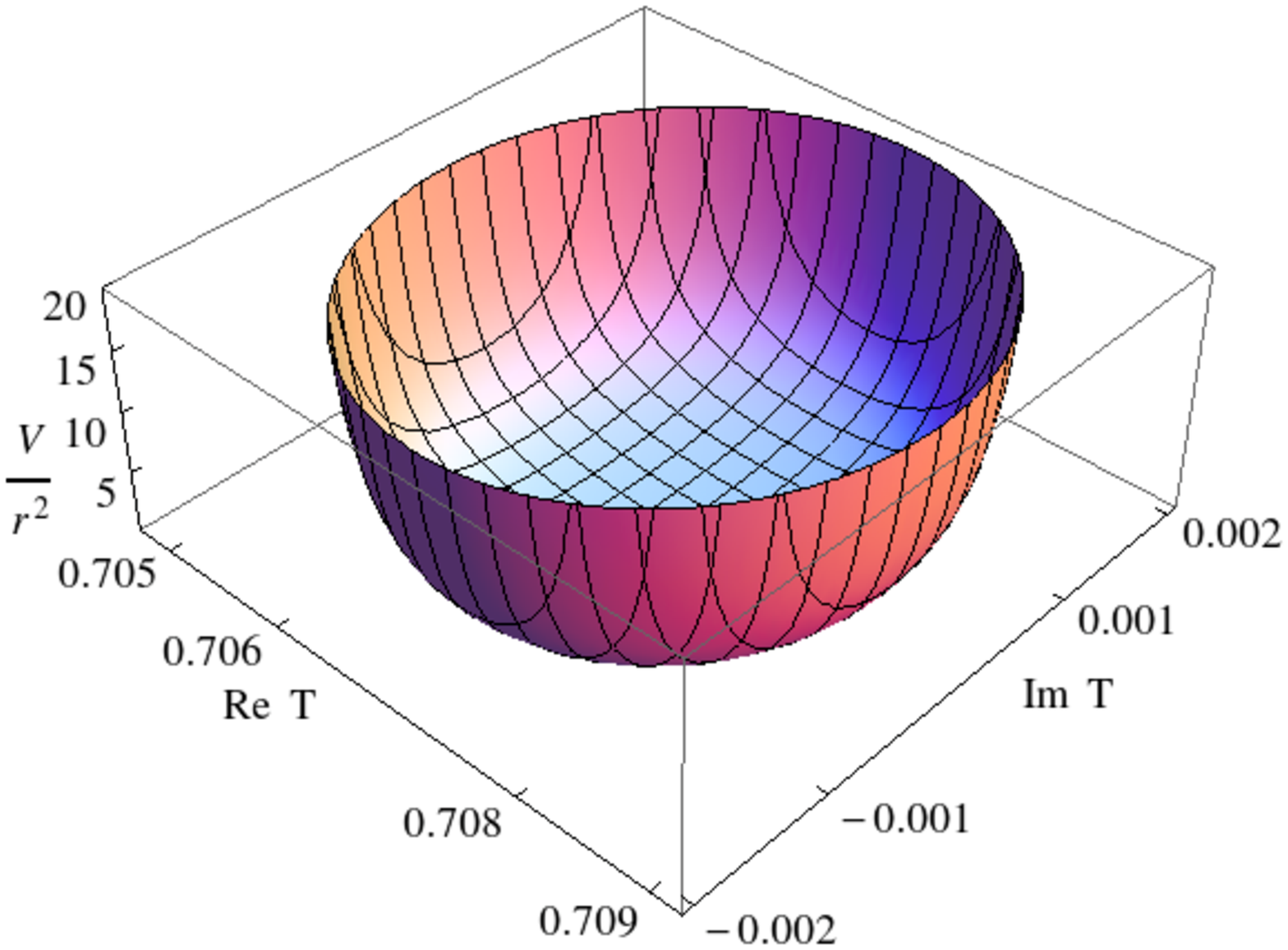}} 
\vspace{-3cm}
\caption{\it The scalar potential for
the real and imaginary components of $T$ at $\phi=0$ in the neighborhood of ${\rm Re}\, T = 1$ 
and ${\rm Im}\, T = 0$ in the model (\protect\ref{modW}, \protect\ref{stable}).}
\label{pot3}
\end{figure}

The shapes of the effective scalar potential in various projections are shown in
Fig.~\ref{pot1}, \ref{pot2} and \ref{pot3}. We see explicitly in Fig.~\ref{pot1} the form of the
effective potential for the real and imaginary components of $\phi$,
assuming that $m= 10^{-5}, \tilde{m}  =10^{-13}$ for $\Lambda=10^{-2}$ the fixed value $2 {\rm Re}\, T = 1$ and ${\rm Im}\, T = 0$.
By construction, the real part of $\phi$ has the desired quadratic potential, and we see that
the effective potential for the imaginary part has a minimum at ${\rm Im}\, \phi = 0$.
Secondly, Fig.~\ref{pot2} shows, correspondingly, that the real parts of $T$ 
and $\phi$ are indeed stabilized in the neighborhood of ${\rm Re}\, T = 1$
and ${\rm Re}\, \phi = 0$. The curvature of the potential for the degree of freedom corresponding to ${\rm Re}\, T$
is difficult to see in this figure, as its mass is ${\cal O}(\tilde{m} /\Lambda)$ in Planck units, 
which is hierarchically smaller than the mass
of the inflaton ${\rm Re}\, \phi$, $m$ in this example.
Thirdly, Fig.~\ref{pot3} shows, correspondingly, that both the real and imaginary parts of $T$ 
are indeed stabilized in the neighborhood of ${2 \rm Re}\, T = 1$ 
and ${\rm Im}\, T = 0$ when $\phi = 0$.

\begin{figure}[h!]
\centering
\vspace{-1cm}
	\scalebox{0.63}{\includegraphics{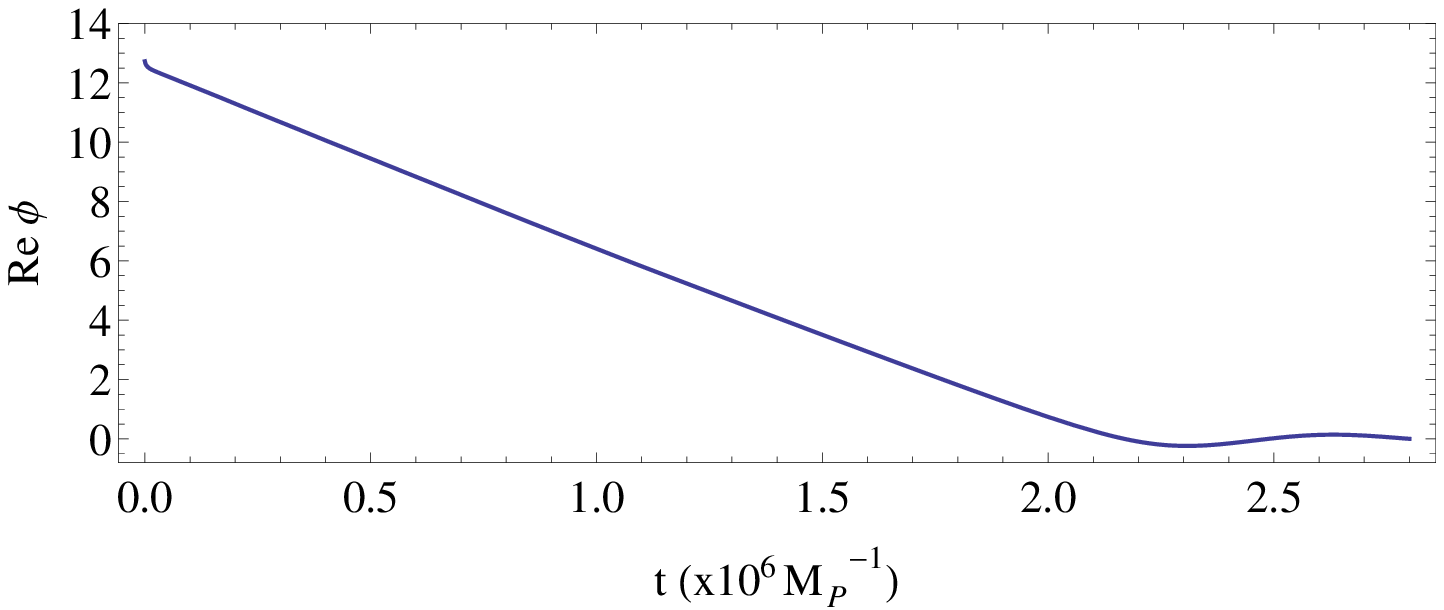}} \\
\scalebox{0.63}{\includegraphics{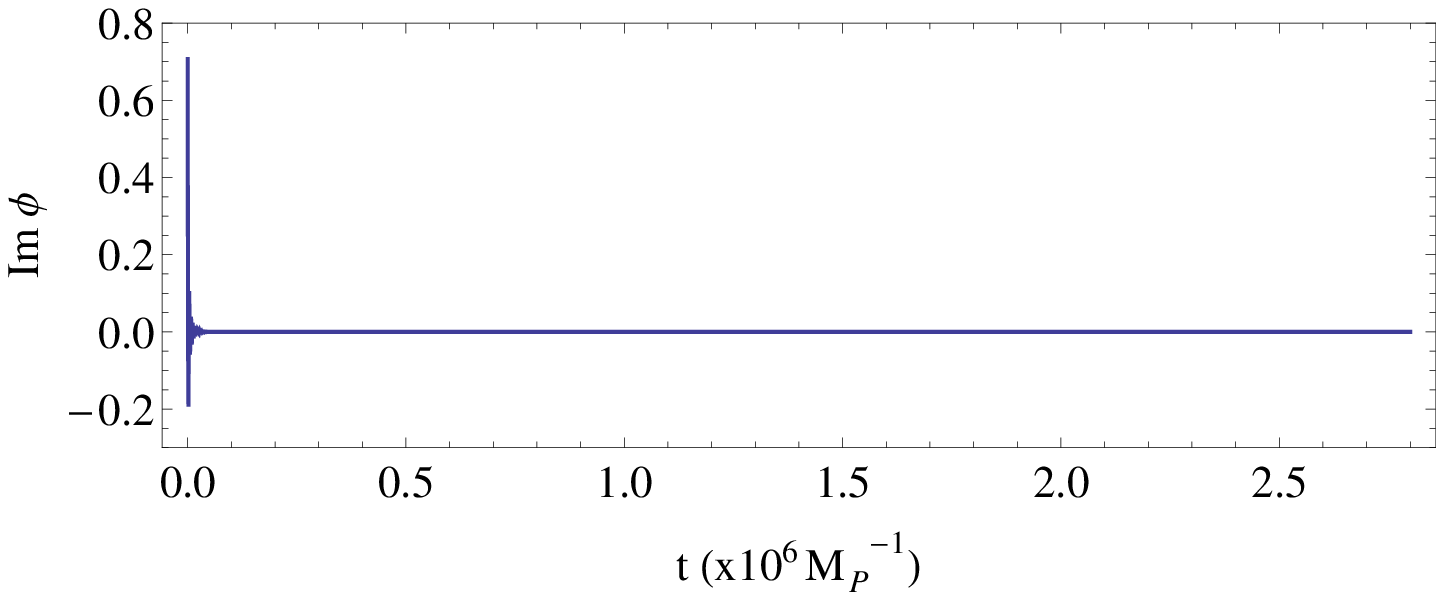}} \\
\scalebox{0.63}{\includegraphics{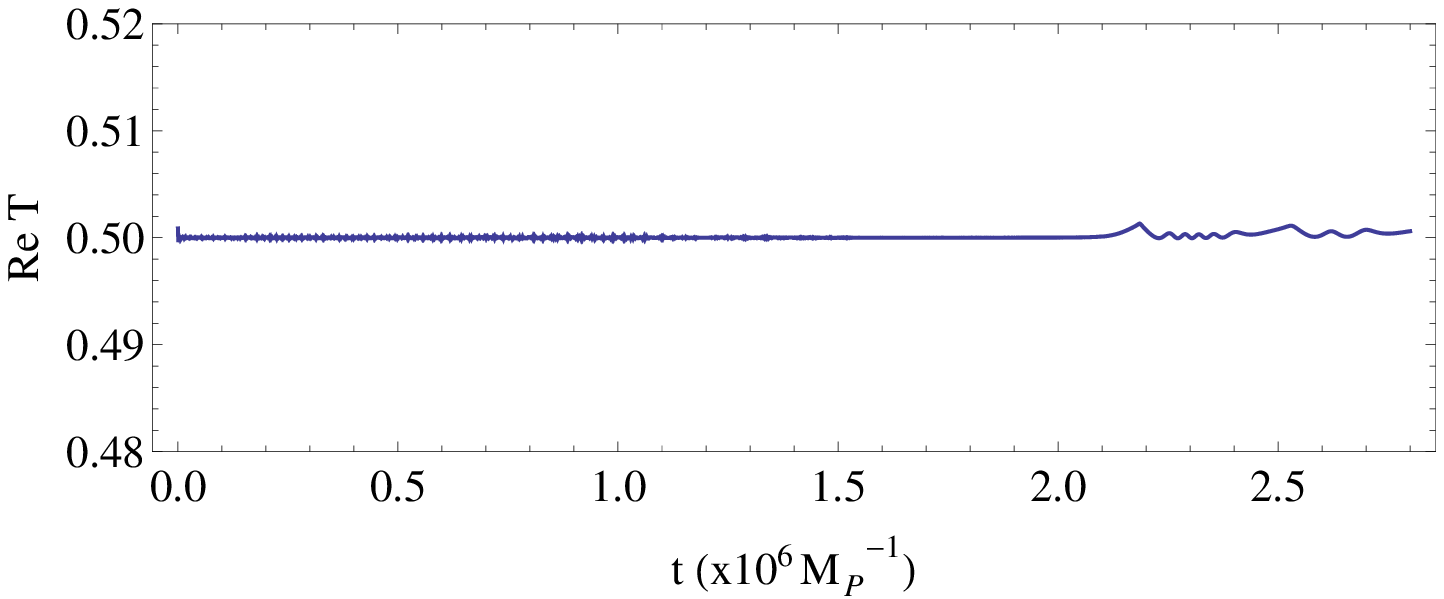}} \\
\scalebox{0.63}{\includegraphics{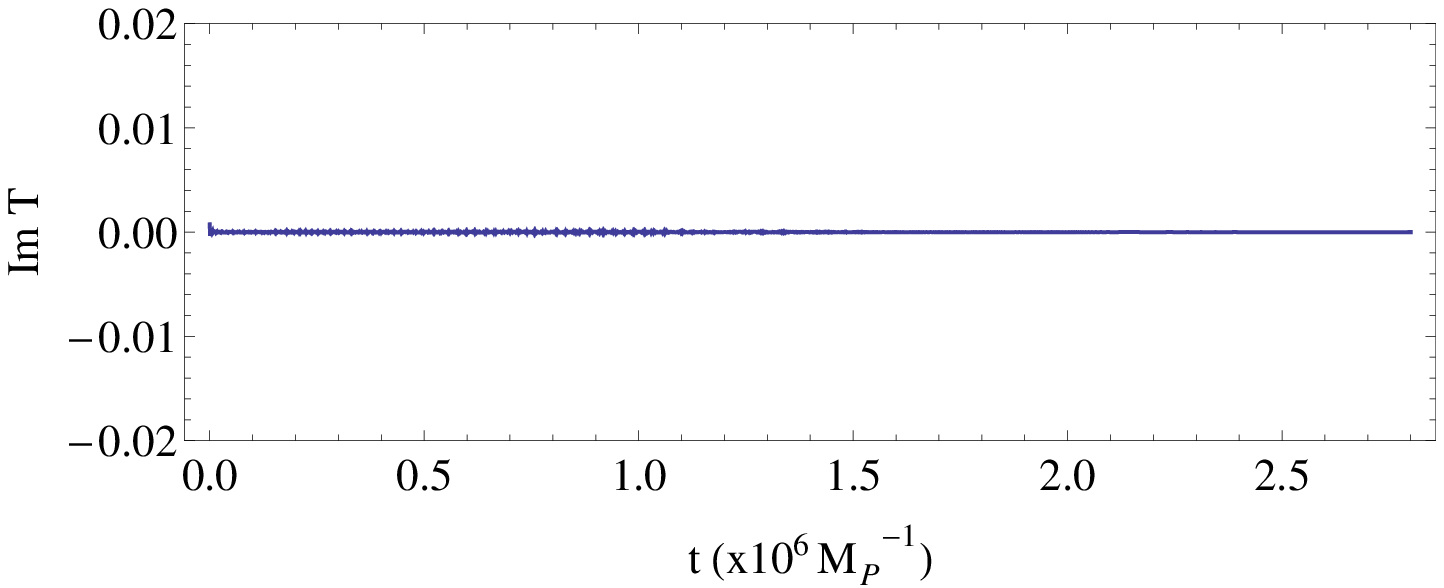}} \\
\scalebox{0.63}{\includegraphics{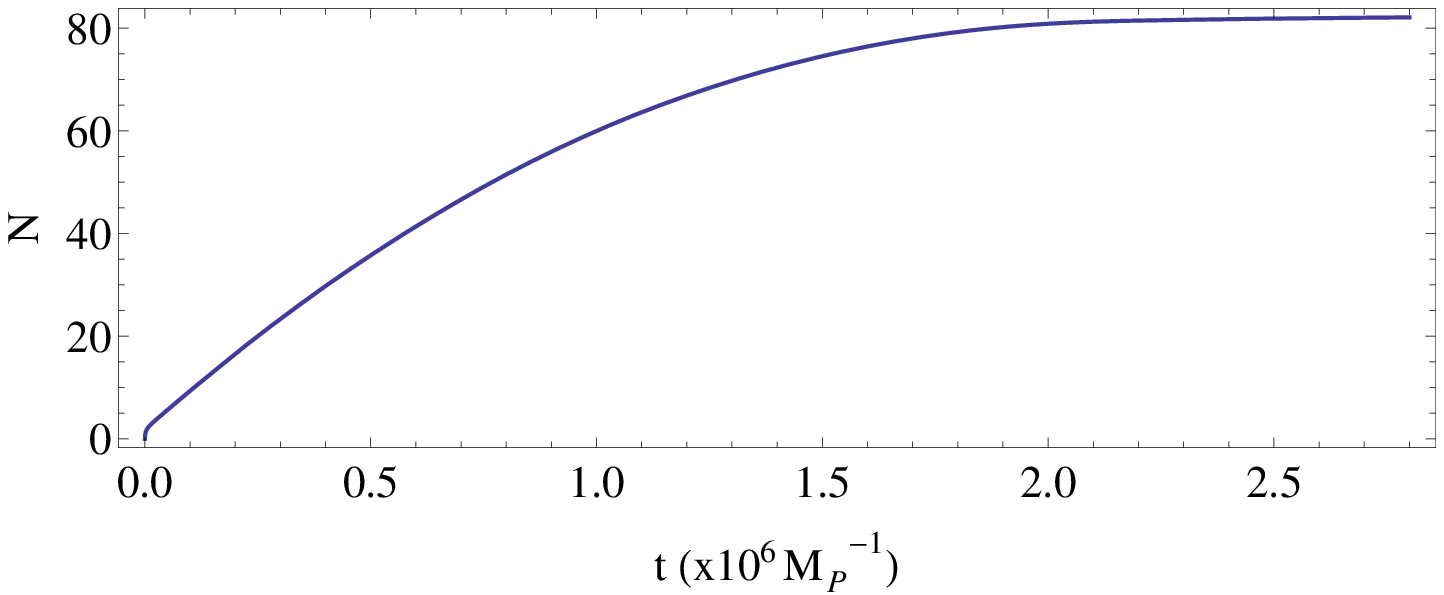}} \\
\caption{\it Analysis of the no-scale quadratic inflationary model given by the K\"ahler potential
(\protect\ref{stable}) and the superpotential (\protect\ref{modW}).
Top panel: Time evolution of the real part of the inflaton $\phi$;
second panel: evolution of the imaginary part of $\phi$;
third panel: evolution of the real part of $T$;
fourth panel: evolution of the imaginary part of $T$;
bottom panel: growth of the  number of e-folds $N$ during inflation.} \label{evolve}
\end{figure}

It is necessary also to verify also that the real and imaginary components of both $T$ and $\phi$
evolve correctly during the inflationary epoch. Accordingly, in Fig.~\ref{evolve} we display
the evolutions of all four components during the inflationary epoch, assuming $d=1$ in
(\ref{stable}), starting from the initial conditions
\beq
\phi_0 \; = \; \frac{1}{\sqrt{2}}(18+i);  \; T_0 \; = \; \frac{1}{\sqrt{2}}(0.7085+0.0012i)
\label{initial}
\eeq
and assuming $\tilde{m} =10^{-13}, m=10^{-5}$ and $\Lambda=10^{-2}$. 
The top, second, third and fourth panels in Fig.~\ref{evolve} display the
evolutions of ${\rm Re}\, \phi, {\rm Im}\, \phi, {\rm Re}\, T$ and ${\rm Im}\, T$, respectively.
We see that the inflaton ${\rm Re}\, \phi$ evolves as expected towards zero, ending with some mild oscillations,
and that there some harmless initial oscillations in ${\rm Im}\, \phi$, while the other field components
remain very close to their values at the minimum of the effective potential throughout the inflationary epoch.
The bottom panel of Fig.~\ref{evolve} displays the evolution of the cosmological scale factor
during the inflationary epoch, demonstrating that a suitable number of e-folds $N$ can be obtained.
The values of the scalar tilt and the tensor-to-scalar ratio are
\beq
(n_s,r)=
\begin{cases}
(0.9596,0.1620) \ \ {\rm for} \, N=50\\(0.9657,0.1429) \ \ {\rm for} \, N=60 \, .
\end{cases}
\label{ourvalues}
\eeq
We conclude that the model (\ref{stable}, \ref{modW}) provides a satisfactory
BICEP2-compatible model of inflation.

\section{Summary and Conclusions}

We have shown that the BICEP2 data on $r$ and the available data on $n_s$ are
consistent (\ref{consistency}) with a simple power-law, monomial, single-field model of inflation,
and that $V = m^2 \phi^2/2$ is the power-law that best fits the available data (\ref{nvalues}).
The required value of $m \simeq 2 \times 10^{13}$~GeV and the small value of the quartic
coupling required for the quadratic potential is to be a good approximation when
$\phi \simeq \sqrt{200} M_{Pl}$ during inflation are technically natural in a supersymmetric model~\cite{ENOT}.
Moreover, it is attractive to identify the inflaton with a singlet (right-handed) sneutrino, since this
value of $m$ lies within the range favoured in Type-I seesaw models of neutrino masses. It is
natural to embed quadratic (sneutrino) inflation within a supergravity framework, and we have
given examples how this may be done in the context of both minimal and no-scale supergravity.

Nevertheless, we would like to reiterate that the BICEP2 measurement of $r$ 
is in tension with the Planck upper limit on $r$,
and emphasize that our choice here to discard the latter and explore the implications of the former
is somewhat arbitrary. In our ignorance, we have no opinion how the tension between the two
experiments will be resolved. If it is resolved in favour of Planck, Starobinsky-like models
would return to favour, which can easily be accommodated in the no-scale supergravity
framework, in particular, with a relatively simple superpotential such as (\ref{oldW}). 
Alternatively, if the resolution favours BICEP2, as we have
shown in this paper, the simplest possible $m^2\phi^2/2$ potential would be favoured,
which offers a very attractive connection to particle physics if the inflaton is identified as a
sneutrino. As we have shown, such a model could also be accommodated within a
no-scale supergravity framework, though at the expense of a more complicated
superpotential such as (\ref{funnyW}) or (\ref{modW}). Models with values of $r$ intermediate between
the ranges favoured by Planck and BICEP2 can also be constructed within the no-scale framework.
A final caveat is that all our analysis is within the slow-roll inflationary paradigm, whereas the
resolution of the tension between Planck and BICEP2 might require going beyond
this framework, e.g., to accommodate large running of the scalar spectral index, a stimulating possibility
that lies beyond the scope of this work.

\section*{Acknowledgements}

The work of J.E. was
supported in part by the London Centre for Terauniverse
Studies (LCTS), using funding from the European
Research Council via the Advanced Investigator Grant
267352. The work of D.V.N. was supported in part by the
DOE grant DE-FG03-95-ER-40917 and would like to thank A.P. for inspiration.
The work of M.A.G.G. and K.A.O. 
was supported in part by DOE grant DE-FG02-94-ER-40823 at the University of Minnesota.


\end{document}